\documentclass[3p,twocolumn]{elsarticle}
\usepackage{diagbox}
\usepackage{lineno,hyperref}
\modulolinenumbers[5]

\journal{Nuclear Instruments and Methods A}

\begin{document}

\begin{frontmatter}

\title{Characterization and performance of the DTAS detector}


\author[ific,fn1]{V.~Guadilla\corref{mycorrespondingauthor}}
\cortext[mycorrespondingauthor]{Corresponding author}
\ead{guadilla@ific.uv.es}
\fntext[fn1]{Current address: Subatech, IMT-Atlantique, Universit\'e de Nantes, CNRS-IN2P3, F-44307, Nantes, France}

\author[ific]{J.L.~Tain}  
\author[ific,debrecen]{A.~Algora}  
\author[ific]{J.~Agramunt}  
\author[jyfl]{J.~\"Ayst\"o}  
\author[nantes]{J.A.~Briz}  
\author[nantes]{A.~Cucoanes}  
\author[jyfl]{T.~Eronen}  
\author[nantes]{M.~Estienne}  
\author[nantes]{M.~Fallot}  
\author[ucm]{L.M.~Fraile}  
\author[istam]{E.~Ganio\u{g}lu}  
\author[ific,surrey]{W.~Gelletly}  
\author[jyfl]{D.~Gorelov}  
\author[jyfl]{J.~Hakala} 
\author[jyfl]{A.~Jokinen}  
\author[ific]{D.~Jordan} 
\author[jyfl]{A.~Kankainen}  
\author[jyfl]{V.~Kolhinen}  
\author[jyfl]{J.~Koponen}  
\author[ipno]{M.~Lebois}  
\author[nantes]{L.~Le Meur}  
\author[ciemat]{T.~Martinez}  
\author[ific]{M.~Monserrate}  
\author[ific]{A.~Montaner-Piz\'a}  
\author[jyfl]{I.~Moore}  
\author[iem]{E.~N\'acher}  
\author[ific]{S.E.A.~Orrigo}  
\author[jyfl]{H.~Penttil\"a}   
\author[jyfl]{I.~Pohjalainen}  
\author[nantes]{A.~Porta}  
\author[jyfl]{J.~Reinikainen}  
\author[jyfl]{M.~Reponen}  
\author[surrey]{S.~Rice}  
\author[jyfl]{S.~Rinta-Antila}  
\author[ific]{B.~Rubio}  
\author[jyfl]{K.~Rytk\"onen}  
\author[nantes]{T.~Shiba}  
\author[jyfl]{V.~Sonnenschein}  
\author[nndc]{A.A.~Sonzogni}  
\author[ific]{E.~Valencia}  
\author[ucm]{V.~Vedia}  
\author[jyfl]{A.~Voss} 
\author[ipno]{J.N.~Wilson}
\author[nantes]{A.-A.~Zakari-Issoufou}

\address[ific]{Instituto de F\'isica Corpuscular, CSIC-Universidad de Valencia, E-46071, Valencia, Spain}
\address[debrecen]{Institute of Nuclear Research of the Hungarian Academy of Sciences, Debrecen H-4026, Hungary}
\address[jyfl]{University of Jyv\"askyl\"a, FIN-40014, Jyv\"askyl\"a, Finland} 
\address[nantes]{Subatech, IMT-Atlantique, Universit\'e de Nantes, CNRS-IN2P3, F-44307, Nantes, France} 
\address[ciemat]{Centro de Investigaciones Energ\'eticas Medioambientales y Tecnol\'ogicas, E-28040, Madrid, Spain} 
\address[ucm]{Universidad Complutense, Grupo de F\'isica Nuclear, CEI Moncloa, E-28040, Madrid, Spain} 
\address[istam]{Department of Physics, Istanbul University, 34134, Istanbul, Turkey} 
\address[surrey]{Department of Physics, University of Surrey, GU2 7XH, Guildford, UK} 
\address[ipno]{Institut de Physique Nucl\`eaire d'Orsay, 91406, Orsay, France} 
\address[iem]{Instituto de Estructura de la Materia, CSIC, E-28006, Madrid, Spain} 
\address[nndc]{NNDC, Brookhaven National Laboratory, Upton, NY 11973-5000, USA}
\begin{abstract}

DTAS is a segmented 
total absorption $\gamma$-ray spectrometer developed for the DESPEC experiment at FAIR. 
It is composed of up to eighteen NaI(Tl) crystals.  In this work we study the performance of this detector 
with laboratory sources and also under real experimental conditions.
 We present a procedure to reconstruct offline the sum of the energy deposited in all the crystals of the spectrometer, 
which is complicated by the effect of NaI(Tl) light-yield non-proportionality.
The use of a system to correct for time variations of the gain in individual detector modules, based on a light pulse generator, is demonstrated.
We describe also an event-based method to evaluate the summing-pileup electronic distortion
in segmented spectrometers. 
All of this allows a careful characterization of the detector with Monte Carlo simulations that is needed to calculate the response function for the analysis of total absorption $\gamma$-ray spectroscopy data. Special attention was paid to the interaction of neutrons with the spectrometer, since they are a source of contamination in studies of $\beta$-delayed neutron emitting nuclei. 

\end{abstract}

\begin{keyword}
$\beta$ decay, total absorption $\gamma$-ray spectrometer, exotic nuclei, NaI(Tl) detector, 
non-proportional
scintillation light yield, Monte Carlo simulations

\end{keyword}

\end{frontmatter}


\section{Introduction}

Decay studies
of exotic nuclear species at the focal plane of the FAIR-NUSTAR Super Fragment Separator in the DESPEC experiment \cite{DESPEC} will provide information on the nuclear structure and the astrophysics  impact of exotic nuclei. Far from stability, the $Q_{\beta}$ values are very large, and the corresponding increase in level density implies, on the one hand, the fragmentation of the $\beta$ feeding into many levels populated in the decay and, on the other hand, the fragmentation of the $\gamma$ intensity between many possible cascades. Total Absorption $\gamma$-Ray Spectroscopy (TAGS) has been shown to be an accurate tool to determine $\beta$-decay intensity distributions for such nuclei far from the valley of $\beta$ stability. This technique avoids the so-called \textit{Pandemonium} effect \cite{Pandemonium}, related to the relatively poor efficiency of 
HPGe detectors. Instead of detecting individual $\gamma$ rays as in high-resolution experiments with HPGe detectors, TAGS aims to detect the full $\beta$-delayed electromagnetic cascade. This is achieved with large scintillator crystals covering a solid angle of $\sim$ 4$\pi$. 

\begin{figure}[h]
\begin{center}
\includegraphics[width=0.37 \textwidth]{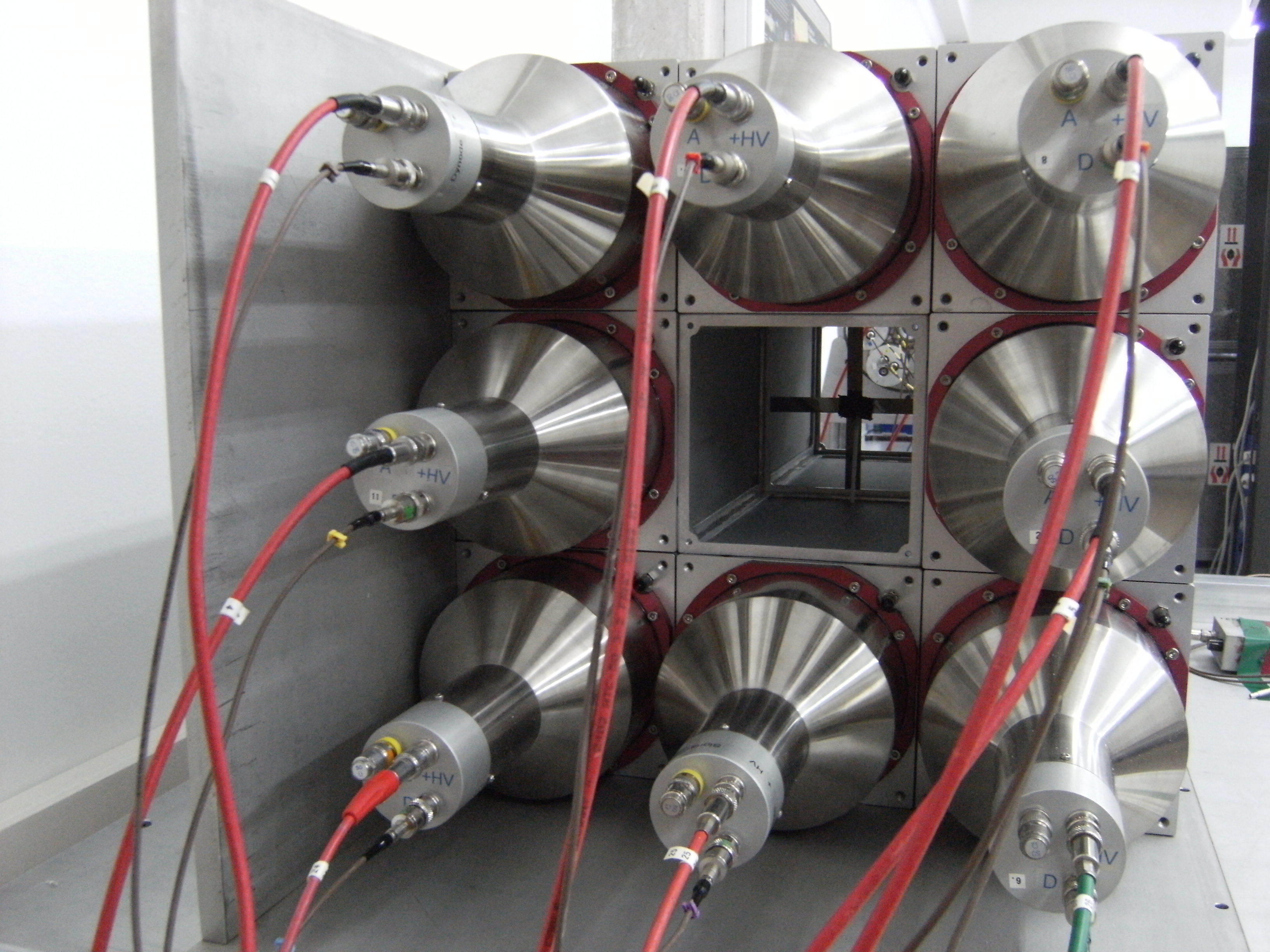} 

\vspace{0.2cm}

\includegraphics[width=0.37 \textwidth]{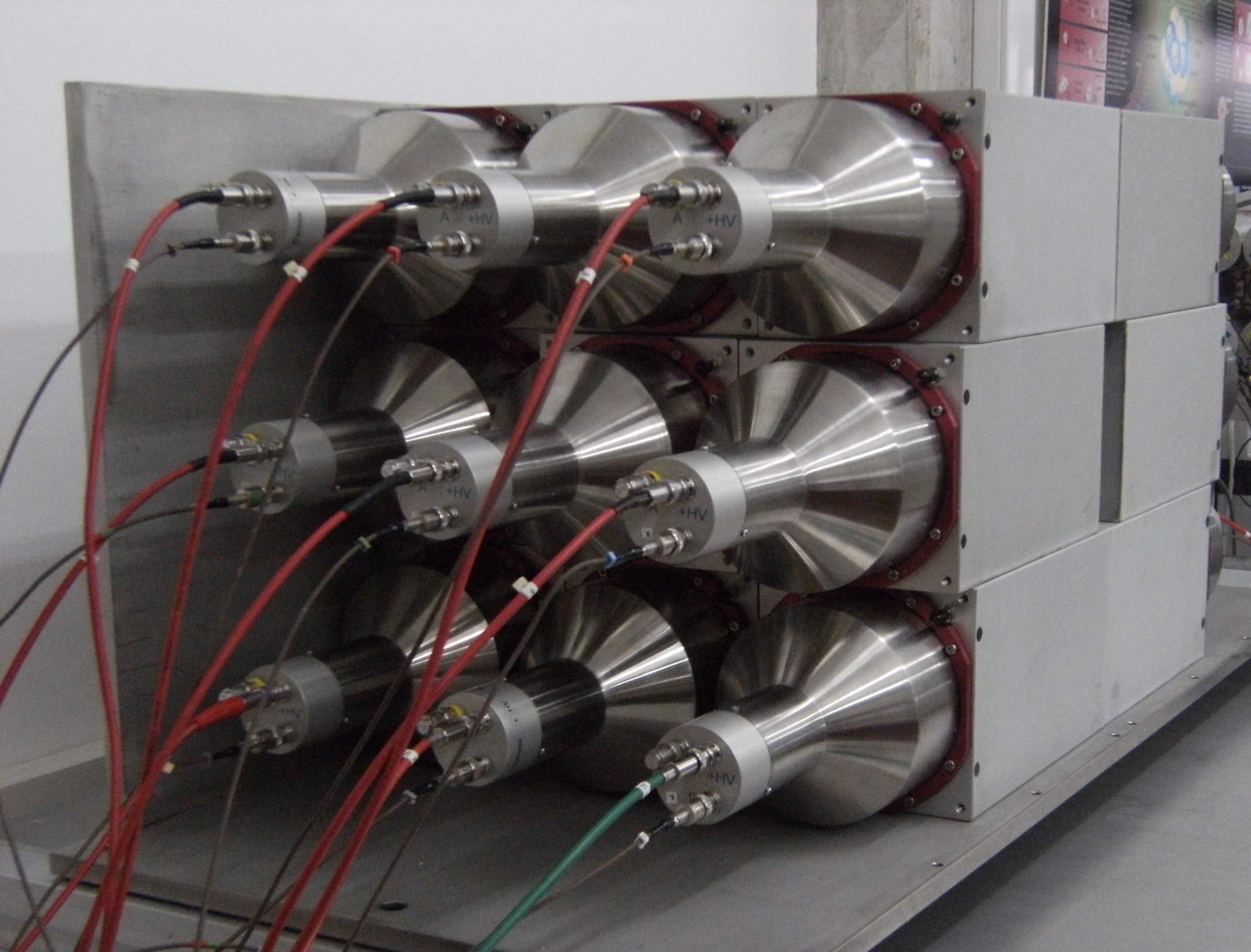} 
\caption{DTAS detector in the sixteen-module configuration (top) and in the eighteen-module configuration (bottom)
without radiation shielding.}
\label{DTAS_conf}
\end{center}
\end{figure}

For this reason, a new spectrometer has been designed and constructed for the DESPEC experiment \cite{DTAS_design}. The Decay Total Absorption $\gamma$-Ray Spectrometer (DTAS) is a segmented detector that consists of a maximum of eighteen NaI(Tl) crystals with dimensions 150~mm $\times$ 150~mm $\times$ 250~mm \cite{DTAS_design}. The advantage of the segmentation in this case is threefold: the possibility to extract information from the multiplicity spectra, as will be explained later, the possibility of using the individual modules as single $\gamma$ detectors, and the mechanical flexibility of the set-up. In fact, we consider two main configurations for DTAS: a sixteen-module configuration designed for 
experiments at fragmentation facilities, 
and an eighteen-module configuration for experiments at ISOL-type facilities.
Both configurations without shielding can be seen in Fig. \ref{DTAS_conf}. In the eighteen-module configuration  
side holes
can be made
by moving away the modules of the horizontal central plane, thus allowing access from both sides of the detector, as shown in Fig. \ref{DTAS_conf} bottom.
In this way DTAS can be combined with ancilliary detectors and it is possible to position a beam pipe in the centre of the spectrometer.
This configuration has recently been commissioned at IGISOL \cite{NIMB_DTAS}, with holes of 10~cm used to place a HPGe detector from one side and the beam pipe with a $\beta$ detector from the other side. The two central modules were separated by 16~cm instead of 10~cm in order to lower their counting rate, so that it was comparable to the 
external modules.
The configuration foreseen for FAIR \cite{DTAS_design}, with sixteen modules, will be coupled to the Advanced Implantation Detector Array (AIDA) \cite{AIDA}. In order to place AIDA in the center of DTAS, the two central modules in the eighteen-module configuration are removed and the two modules above the central hole are supported by a specially designed aluminium frame with external dimensions identical to a module, as shown in Fig. \ref{DTAS_conf} upper panel.

The shielding surrounding DTAS is composed of stainless steel sheets, lead bricks and aluminium, and it served to reduce the background counting rate by one order-of-magnitude in the measurements of this work. The 
allocation of individual modules to positions in the arrangement was done
according to their resolutions, ranging from 7$\%$ to 9$\%$
at 661.7~keV, so that the positions associated with the lowest counting rates (the eight corners of the assembly shown in Fig. \ref{DTAS_conf}) were occupied by the modules with the poorest resolution.

The outline of the article is the following: in section \ref{sec-1} we will describe the procedure to reconstruct the full energy deposited in the detector from the signals of the individual modules. In section \ref{sec-2} a method to evaluate the summing-pileup contamination will be explained, and its validation with calibration sources will be discussed. Finally, the Monte Carlo (MC) response function of the detector will be described in section \ref{MC_DTAS}, and the reproduction of several calibration sources and the neutron contamination coming from $\beta$-delayed neutron emitters will be 
discussed.

\section{Total energy reconstruction: hardware sum and software sum}\label{sec-1}

In this section we will describe the electronic chain employed to process the signals from the individual modules of DTAS, and the procedure to reconstruct the total energy deposited in the detector. In particular, two methods to calculate the total 
energy sum
will be 
discussed:
the hardware sum and the software sum. 
 
\subsection{Signal processing}\label{electronics}

In order to analyse data from DTAS we have to reconstruct accurately, for each event, the energy deposited in the full spectrometer and its multiplicity, $M_m$ (number of modules that fire above the threshold). The full energy released in the spectrometer is obtained by summing the energy deposited in the individual modules, either electronically or via software. The electronic chain to process the signals from the modules was designed with this idea in mind, and it is represented in Fig. \ref{electronics}. 

\begin{figure*}[h]
\begin{center} 
\includegraphics[width=1 \textwidth]{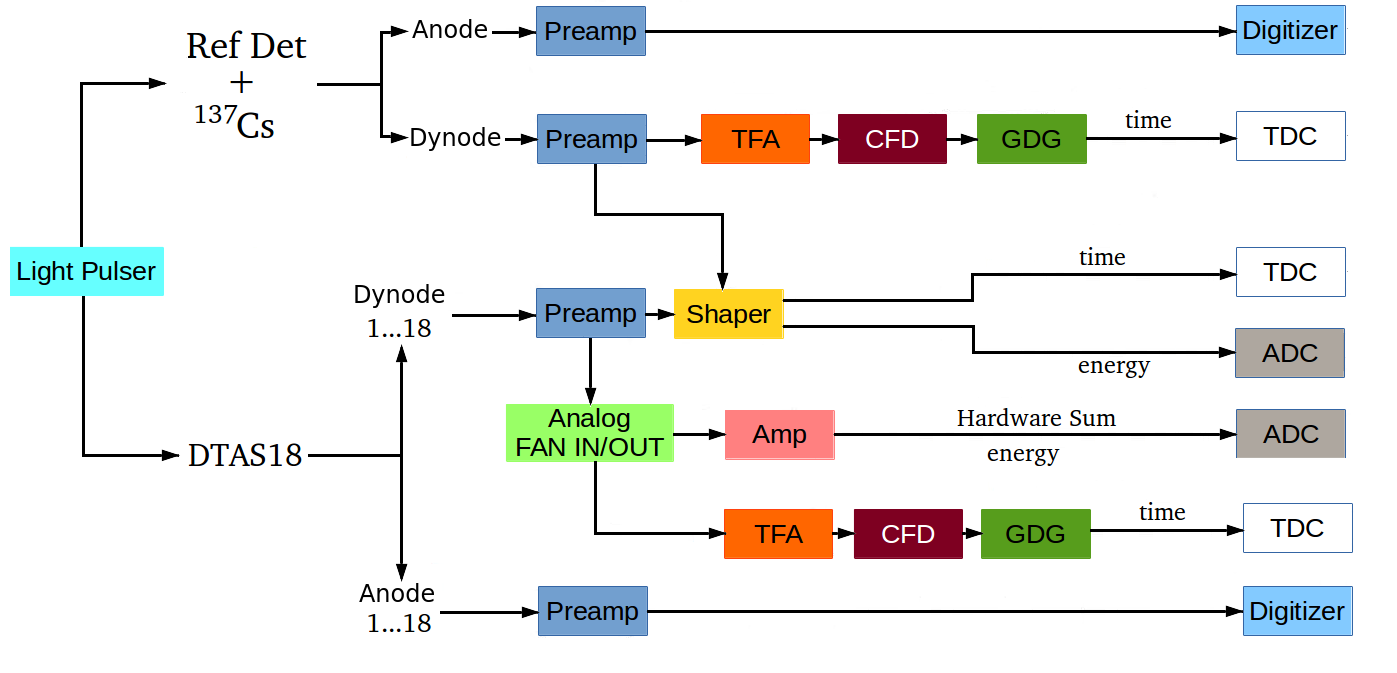}
\caption{Schematic diagram of the electronic chain. The labels correspond to: Preamplifier (Preamp), Spectroscopic Amplifier (Amp), Timing Filter Amplifier (TFA), Constant Fraction Discriminator (CFD), Gate/Delay Generator (GDG), Time to Digital Converter (TDC), Analog to Digital Converter (ADC).}
\label{electronics}
\end{center}
\end{figure*}

We use Mesytec MSI-8p preamplifiers \cite{mesytec} for both  
anode and dynode signals from the photomultiplier tubes (PMTs).
After the preamplifier, dynode
signals
are split into two branches; one branch is sent to a CAEN N625 Quad Linear FAN-in FAN-out \cite{caen}, and the other to Mesytec MSCF-16 
shapers. 
The N625 module acts as an analog signal adder
and one of the outgoing signals is processed in an ORTEC 671 amplifier \cite{ortec} to produce the sum energy signal
(hardware sum)
sent to the analog to digital converter (ADC), a CAEN V785 module, 
of the data acquisition system (DACQ). 
Another output from the N625 module is used to construct a 
common stop signal sent to a time to digital converter (TDC), CAEN V775, using an ORTEC 474 Timing Filter Amplifier and an
ORTEC 584 Constant Fraction Discriminator. 
The MSCF-16 shapers provide individual energy and timing output signals that are sent to the individual channels of the ADC
and TDC modules respectively.
The anode signals after the preamplifier are sent to sampling digitizers of a second digital DACQ, 
running in self-triggered mode, which is not discussed in this publication.

In order to carry out 
the hardware
sum properly we need to match the gains of 
the different PMTs
by adjusting the high voltage (HV) applied to them, so that 
the signals of individual modules
are aligned. Note that aligned here means having the same amplitude for the same 
energy deposited.

The software sum is reconstructed offline from the individual signals processed with the 
MSCF-16
shapers. In the following subsections we will show a method of correcting possible changes in the gain of the modules, as well as the way to perform properly the alignment and determine the software sum of these signals.

\subsection{Gain correction system}\label{gain}

\begin{figure}[h]
\begin{center}
\includegraphics[width=0.4 \textwidth]{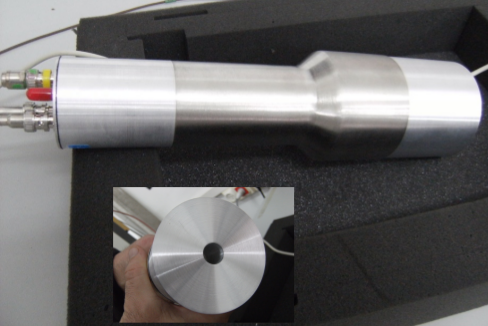} 
\caption{NaI(Tl) reference well type detector. Inset: a view of the front face with the hole where the weak $^{137}$Cs source is placed.}
\label{Ref_det}
\end{center}
\end{figure}

A system to correct changes in the gain of 
individual
modules has been developed. These changes may be due to temperature 
variations
\cite{Temperature_NaI}, drift of the PMT
current
and fluctuations in the HV supply. 
In this system the gain of each module is monitored
checking the position of 
the peak produced by a pulsed light source.
An additional external reference detector, 
with a weak $^{137}$Cs radioactive source, 
is used to monitor the stability of the light pulse generator. The following elements are employed in this system:

\begin{itemize}
\item An external reference well-type NaI(Tl) detector of 3'' diameter $\times$ 3'' length manufactured by Saint Gobain \cite{saint-gobain}. The well has 15~mm diameter and 40~mm depth. The 
crystal is
mounted on a 3'' diameter ETI 9305 PMT as shown in Fig. \ref{Ref_det}.

\item A 490~nm light pulse generator model 6010 from BNC \cite{bnc}. 
The generator is triggered with an external 100~Hz clock signal.

\item A 2~m long bundle of borosilicate glass fibres split into 20 bundles of 2~mm diameter, manufactured by FiberTech Optica \cite{fibertech}. The fibres are terminated with SMA type connectors.

\item A weak $^{137}$Cs source of $\sim$ 300~Bq. 

\end{itemize}

\begin{figure}[h]
\begin{center}
\includegraphics[width=0.5 \textwidth]{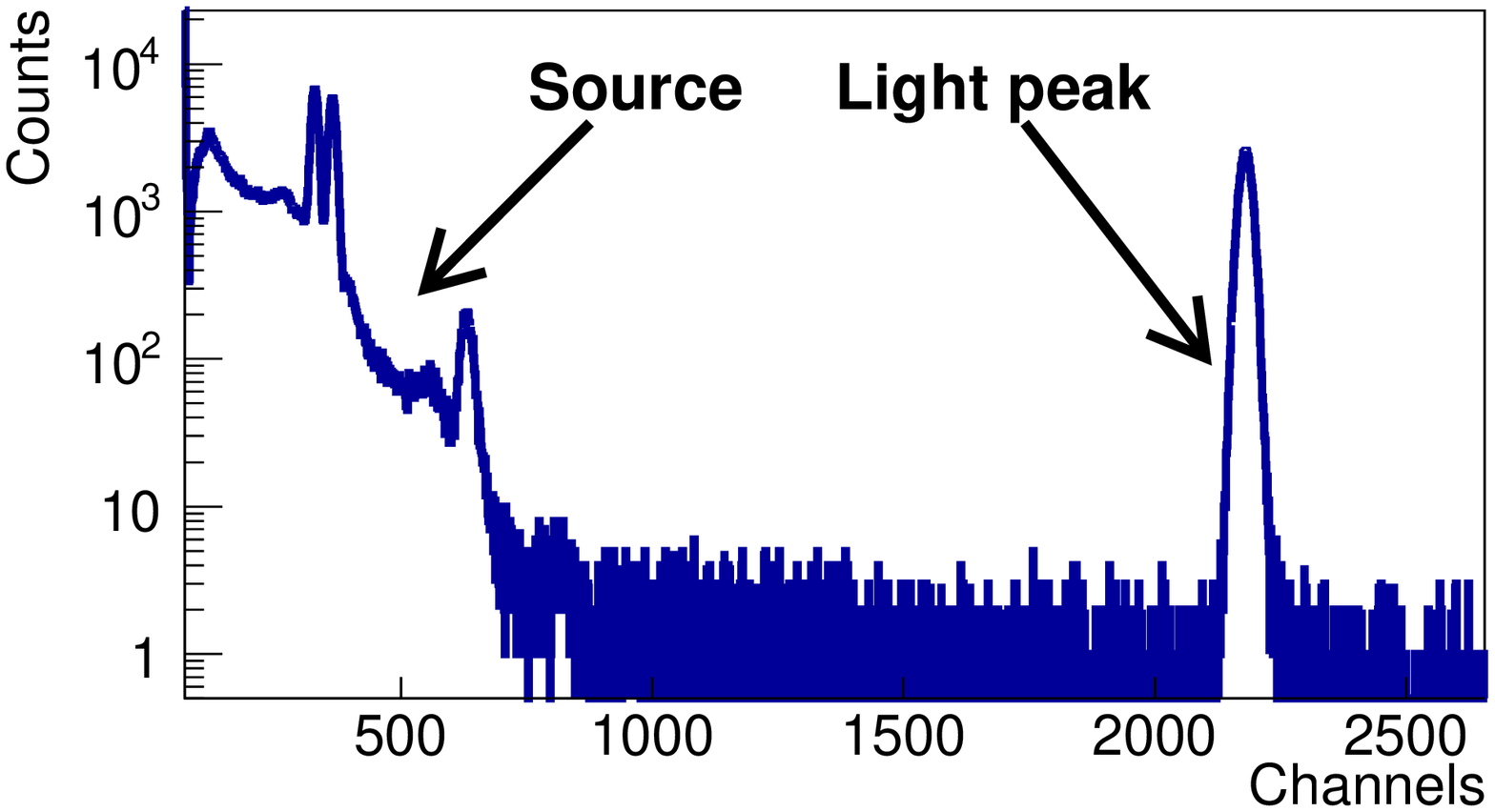} 
\vspace{0.2cm}
\includegraphics[width=0.5 \textwidth]{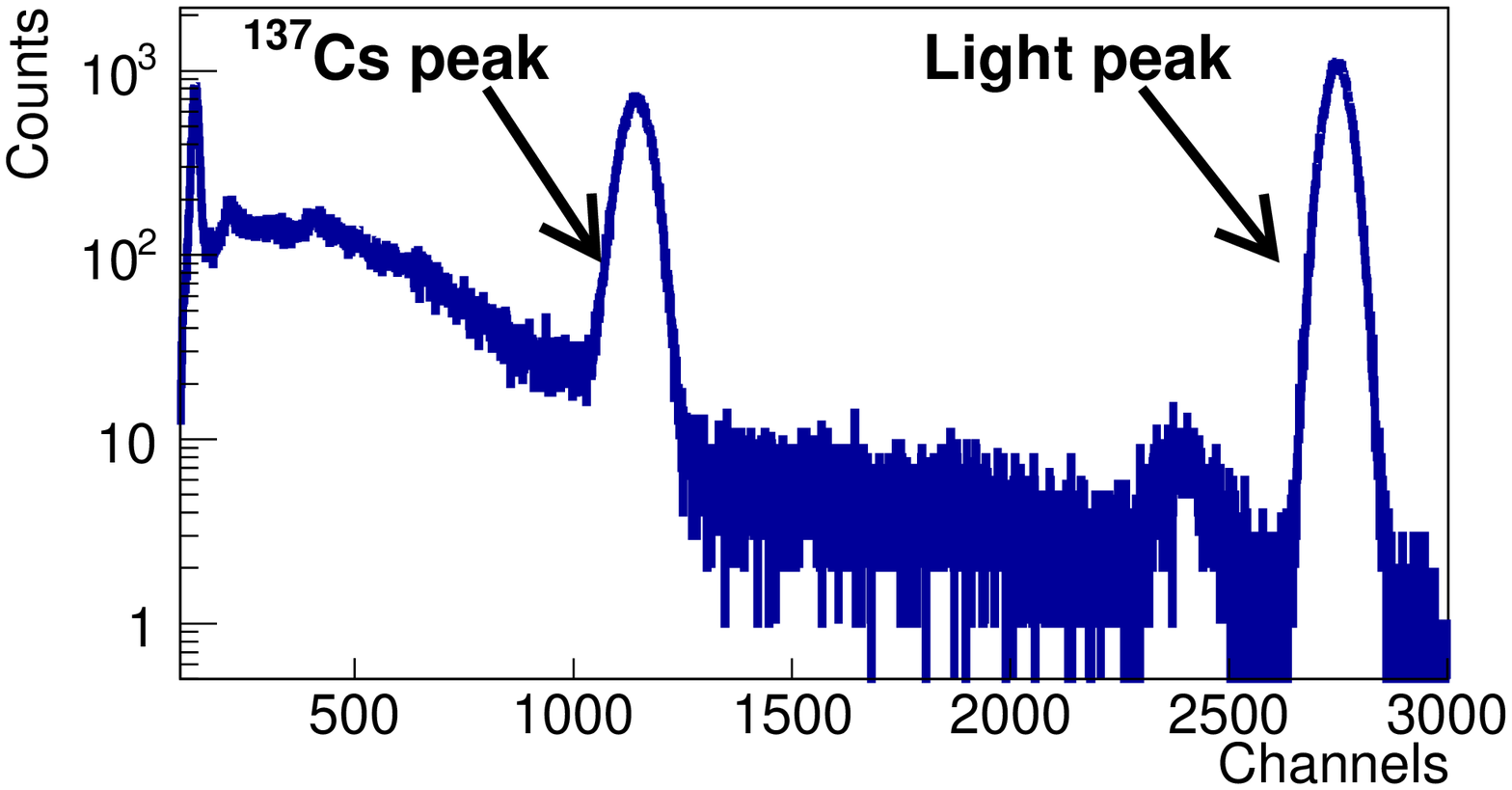} 
\caption{Individual DTAS detector spectrum with the light pulser peak in a $^{60}$Co measurement (top). Reference detector spectrum with the 661.7~keV peak from the weak $^{137}$Cs source, and the light pulser peak (bottom).}
\label{Reference_peaks}
\end{center}
\end{figure}

The fibre bundle 
splitter
 is used to distribute the light pulse from the generator to the reference detector and to each of the eighteen modules. The $^{137}$Cs source is placed inside the well of the reference detector. The reference detector is surrounded by lead shielding and is placed close to DTAS. Since both the reference detector and DTAS have shielding, this weak source does not affect the DTAS measurements. 
The position of the 661.7~keV peak in the well detector provides a reference for possible
changes in the gain of this detector. Comparing the position of the light pulser peak 
with this peak we can determine if there are variations of the intensity of the light source.
With this information we can separate in each module variations in the gain from variations
in the light source intensity.
The 
gain
correction is calculated for short 
time intervals,
and the procedure will be detailed in the next subsection. An example of the spectra of the reference detector and 
one
individual module of DTAS showing the light pulser peaks can be seen in Fig. \ref{Reference_peaks}. 

In order not to disturb the measured individual spectra, the 
peak due to the light pulser has to be located beyond the energy region of interest, 
see Fig. \ref{Soft_sum_ex} as an example. 
When choosing an optical fibre bundle for each module, we took into account that each of the 20 
bundles
does not transport the same amount of light, and the individual modules do not convert the same amount of incident light into the same signal amplitude in the PMT. For both reasons, in order to 
minimize the difference in position of light pulser peaks between modules
we assigned the 
bundles 
that transport more light with the worst modules in terms of light conversion. 

Apart from applying the gain correction offline, the gain correction system could also be used for maintaining the alignment of the signals of the modules 
during the measurement
by applying 
periodic
HV corrections to the PMTs. This 
requires information about the dependence of the gain with the HV for each module. Although we have tested this online correction 
method, it was not used in the actual measurements.

\subsection{Software sum}\label{sum}

Just as in the case of the hardware sum, before performing the software sum the amplitude of the signals 
stored for each event must be properly aligned.
Although 
signal
amplitudes were 
gain-matched
before the FAN-in FAN-out for the hardware sum, and even though the gains of the shapers are set to a common value, 
the stored amplitude information needs to be corrected
due to slight variations in gain and offset of the individual electronic channels. 

The first idea for 
making
this alignment was to convert signal amplitude (proportional to light collected) into energy for each of the individual channels. This conversion between light collected and deposited energy is what we will call energy calibration. A solution like this has been successfully adopted for a 12-fold segmented BaF$_2$ spectrometer in previous works \cite{vTAS_PRL,Zak_PRL,vTAS_PRC,Simon_PRC}. Nevertheless, we soon realized that it can not be applied in the case of a segmented detector made of NaI(Tl) because of the non-proportionality of the light yield in this material \cite{Non_Prop_Discover, non-prop3}. The reason is 
related to
what was 
pointed out in \cite{TAS_MC}, 
explaining the shift of
the position of full energy peaks due to $\gamma$-ray cascades 
with respect to single $\gamma$-ray peaks of the same energy. 
For every primary electron created by $\gamma$-ray interactions there is a shift of about
10~keV in the apparent energy. Since $\gamma$-rays of several hundreds of keV to a few MeV
typically require of the order of three interactions (two Compton, one photoelectric) to deposit the full energy
this explains why
for a $\gamma$-cascade of two $\gamma$-rays ($\gamma$-multiplicity, $M_{\gamma}$=2) the shift is approximately 30~keV, while for $M_{\gamma}$=3 it is 60~keV and so on. In the case of a segmented detector the situation is more complicated, and the shift depends not only on the $\gamma$-multiplicity, $M_{\gamma}$, but also on the number of modules where the energy is deposited, $M_m$,
which determines the distribution of the number of primary electrons in each module.
Taking into account the different ways that $3 \times M_{\gamma}$ electrons
can be distributed in $M_{m}$ modules one can determine that
the apparent energy shifts follow approximately the numbers in Table \ref{non-prop_TABLE}.
The first row in the table corresponds to the behaviour of a single NaI(Tl) crystal spectrometer like LUCRECIA at ISOLDE \cite{LucreciaTAS} or the LBNL spectrometer used at GSI \cite{NIMB_TAS_GSI}.

\begin{table}[h]
\begin{center}
\begin{tabular}{ccccc}
\textbf{$M_{\gamma}$} &\makebox[3em]{1}&\makebox[3em]{2}&\makebox[3em]{3}
&\makebox[3em]{4}\\
\hline
\textbf{$M_m$} &&&&\\
$1$ & $0$ & $+30$ & $+60$ & $+90$\\
$2$ & $-30$ & $0$ & $+30$ & $+60$\\
$3$ & $-60$ & $-30$ & $0$ & $+30$\\
$4$ & $-90$ & $-60$ & $-30$ & $0$\\
\hline
\end{tabular}
\caption{Shift in keV of the sum peak
position 
due to the non-proportionality of the light yield in a segmented NaI(Tl) spectrometer 
when
the individual modules 
are
calibrated in energy before the software sum.}
\label{non-prop_TABLE}
\end{center}
\end{table}

\begin{figure}[h]
\begin{center}
\includegraphics[width=0.5 \textwidth]{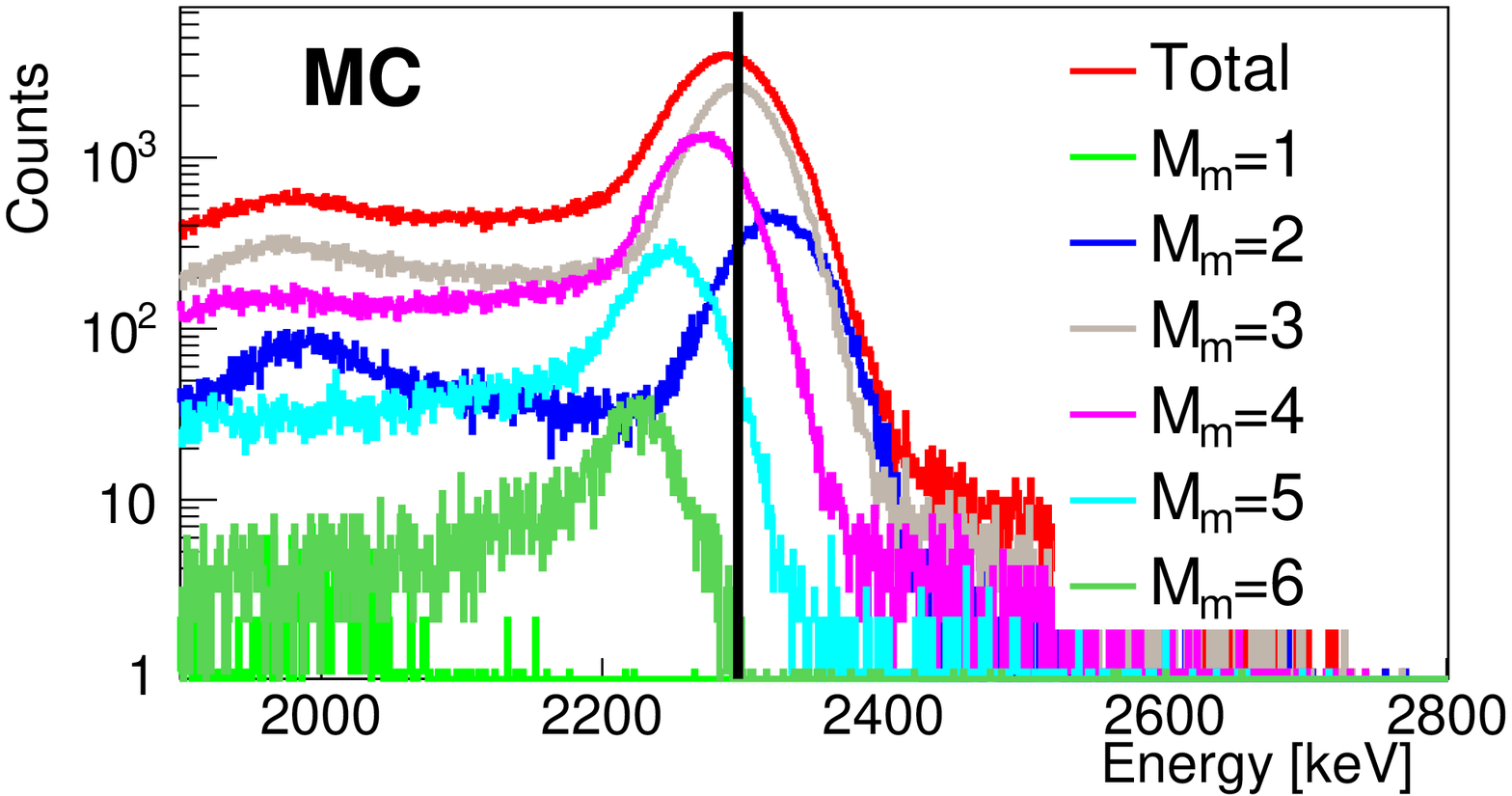} 

\vspace{0.2cm}

\includegraphics[width=0.5 \textwidth]{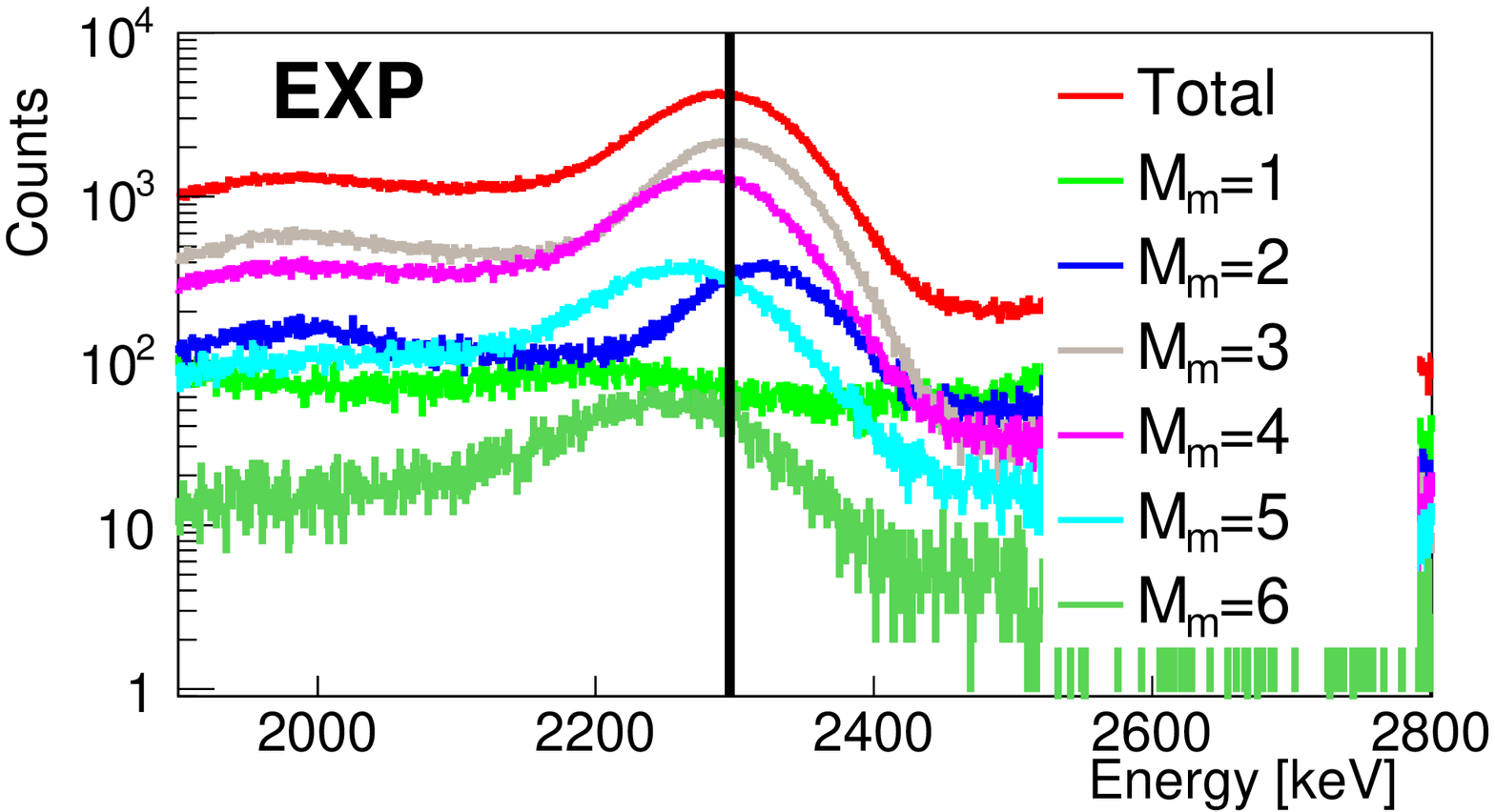} 
\caption{$^{22}$Na software sum of the light produced in the individual detectors 
calibrated
in energy. Both the MC (top) and the experimental measurement (bottom) show the shifts from Table \ref{non-prop_TABLE}. The vertical black line corresponds to the sum of the energies of the three $\gamma$-rays involved. (For interpretation of the references to color in this figure caption, the reader is referred to the web version of this paper.)}
\label{Shift_22Na}
\end{center}
\end{figure}

For a single crystal spectrometer the non-proportionality is not a problem as far as this effect is included in the MC simulations in the way detailed in \cite{TAS_MC}. Likewise, it does not present any problem for the hardware sum of a segmented NaI(Tl) spectrometer, 
as long as the PMTs are gain-matched.
However, the consequence of 
applying an energy calibration to individual modules before summing, 
is that the resolution of the sum peaks is worsened due to the displacement of the different multiplicities contributing to the sum. The non-proportionality of the light yield in NaI(Tl) is known to have an important contribution to the resolution of single crystal detectors \cite{non-prop1,non-prop2}, but this is an additional effect for multi-crystal detectors. In Fig. \ref{Shift_22Na} these shifts are shown for a measurement of $^{22}$Na ($M_{\gamma}$=3) and for the corresponding MC simulation of this source that includes the non-proportionality of the light yield as in \cite{TAS_MC}. In both cases an energy calibration has been applied to all the individual modules before summing. The vertical black line corresponds to 2296.5~keV, the sum of the energies of the three $\gamma$-rays involved: 511~keV, 511~keV and 1274.5~keV. The sum peaks of the different multiplicities are not aligned, showing a displacement in agreement with Table \ref{non-prop_TABLE}. Only $M_m$=3 is aligned with the nominal sum, since it corresponds to a 0~keV shift in Table \ref{non-prop_TABLE}, with three $\gamma$-rays detected in three crystals. Note that the experimental spectra are not background subtracted, whereas the MC is only widened by the light function from \citep{TAS_MC}, without taking into account 
additional contributions to the resolution.

In summary, in order 
to maintain
the resolution, we have to align the 
stored
amplitudes of the individual modules, thus reproducing with our software sum the same behaviour of the hardware sum, and, equivalently, of a single crystal detector. 
In addition there are other effects that may worsen the resolution, like changes in the gain of the PMTs and the electronic chain. The correction we have to apply to counteract these effects is applied sequentially on a reduced number of events. The number of events should be sufficiently large to determine the peak positions accurately and sufficiently small to limit the effect of gain variations during the acquisition time.  We have verified that 1 million events (that corresponds to approximately 4 minutes for a typical counting rate of 4-5~kHz in DTAS), fulfill this condition.

The stored amplitude is represented by the bin number in the histogram accumulated by each ADC 
channel (detector module) and the first step is to determine the offset and the gain. To determine the ADC offset for each channel we use the position of the peak due to the electronic noise. The gain is obtained from the position of the two peaks from a calibration run with a $^{22}$Na source (511~keV and 1274.5~keV). With the offset and gain so obtained the alignment 
of the first million events is performed choosing one arbitrary module as a reference.



After the alignment, the reference values of the parameters involved in the gain correction procedure are determined. 
The ADC offset is represented by $a_j$, with $j=0...18$, with $j=0$ being the well detector and $j=1...18$ the DTAS modules. 
The reference position of the light pulser peak for each module, $L_j$, is 
obtained by peak fitting.
Analogously, the $^{137}$Cs peak and the light pulser peak reference positions for the well detector, $P_0$ and $L_0$ respectively,  are determined.

The next group of one million events is then processed.
We define $L^{'}_j$ as the new light pulser peak position of module $j$, $b_j$ as the gain change factor of module $j$, and $C$ as the change factor in the light source intensity. 
The procedure described below is followed in order to calculate the gain corrections and sum the amplitudes of all modules stored in each event:

\begin{itemize}

\item The new position of the $^{137}$Cs peak for the 
well detector, $P^{'}_0$, is determined, as well as the position of the light pulser peak $L^{'}_0$.

\item The change in the gain of the PMT of the 
well detector is calculated:

\begin{center}
\begin{equation}
b_0=\frac{P_0-a_0}{P^{'}_0-a_0}
\end{equation}
\end{center}

\item The 
change in the light produced by the light pulse generator, $C$, is calculated:

\begin{center}
\begin{equation}\label{Lfluc}
C=\frac{L_0-a_0}{b_0(L{'}_0-a_0)}
\end{equation}
\end{center}

\item $L^{'}_j$ is determined for each of the DTAS modules, and with this value the gain change 
factor
is calculated taking into account the change of intensity of the light source from Equation \ref{Lfluc}:

\begin{center}
\begin{equation}
b_j=\frac{L_j-a_j}{C(L^{'}_j-a_j)}
\end{equation}
\end{center}

\end{itemize}

\begin{figure}[!h]
\begin{center}
\includegraphics[width=0.5 \textwidth]{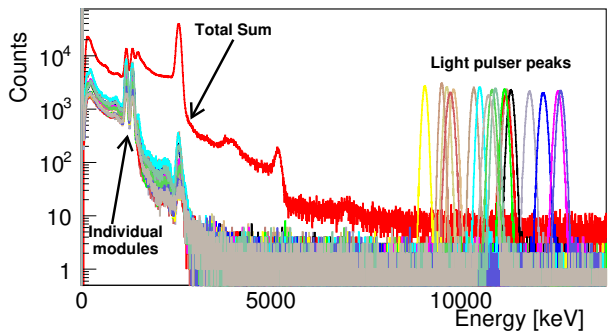}  
\caption{Software sum spectrum of a $^{60}$Co source 
measurement and individual module spectra showing  the alignment. The peaks in individual modules 
above 8~MeV are due to the light pulser.}
\label{Soft_sum_ex}
\end{center}
\end{figure}

Once the parameters $b_j$ are determined we reprocess the same group of events applying the
gain correction factor in order to align the amplitudes of all modules to the first group of events used 
as a reference.
As a result of applying this procedure, the software sum 
can be performed properly,
as can be seen in Figure \ref{Soft_sum_ex} for a $^{60}$Co source. At this point an energy calibration can be applied (a conversion between light collected and energy) by using single peaks ($M_{\gamma}$=1), as in the case of the hardware sum.

\begin{figure}[!h]
\begin{center} 
\includegraphics[width=0.5 \textwidth]{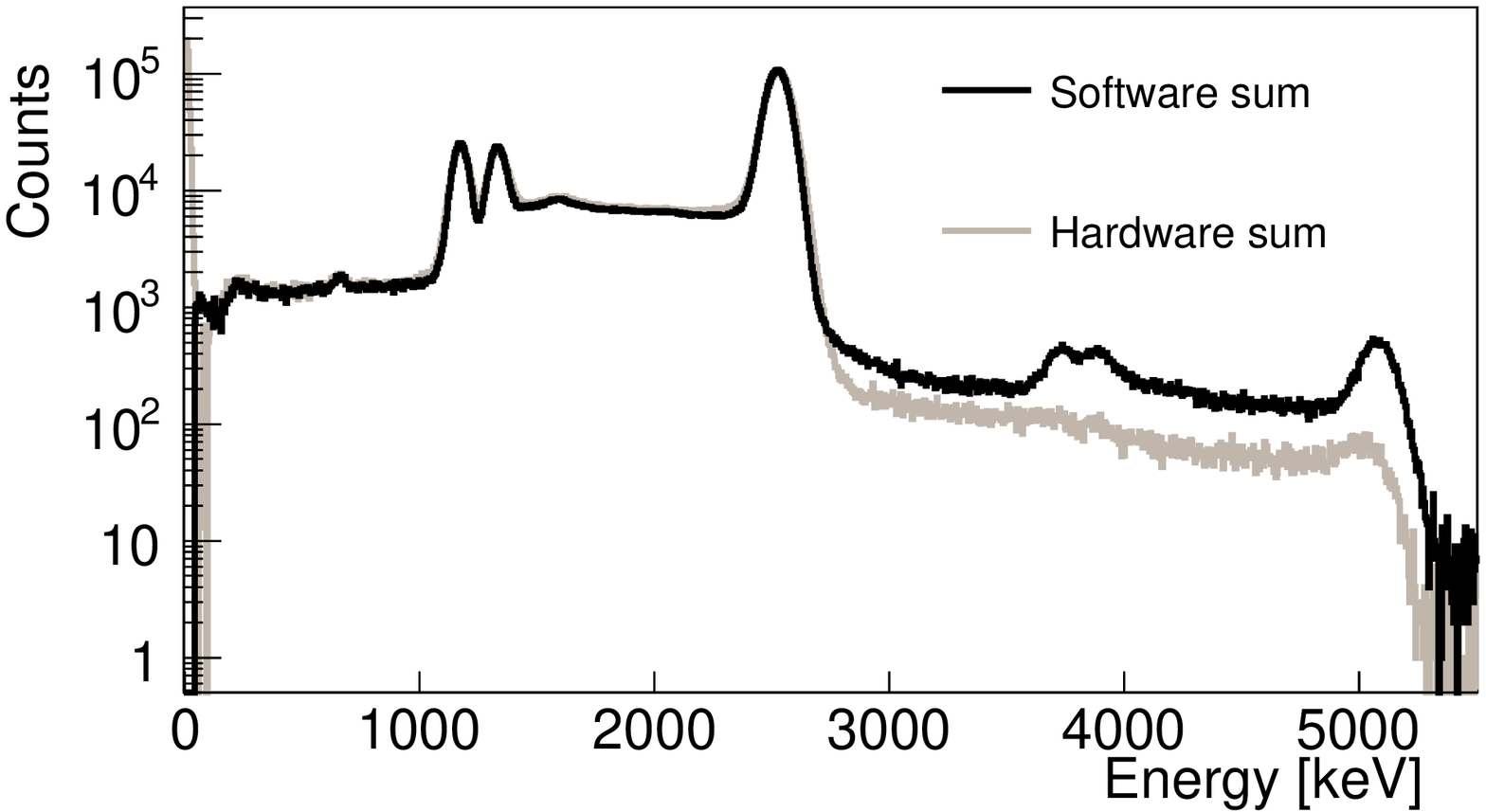} 
\includegraphics[width=0.5 \textwidth]{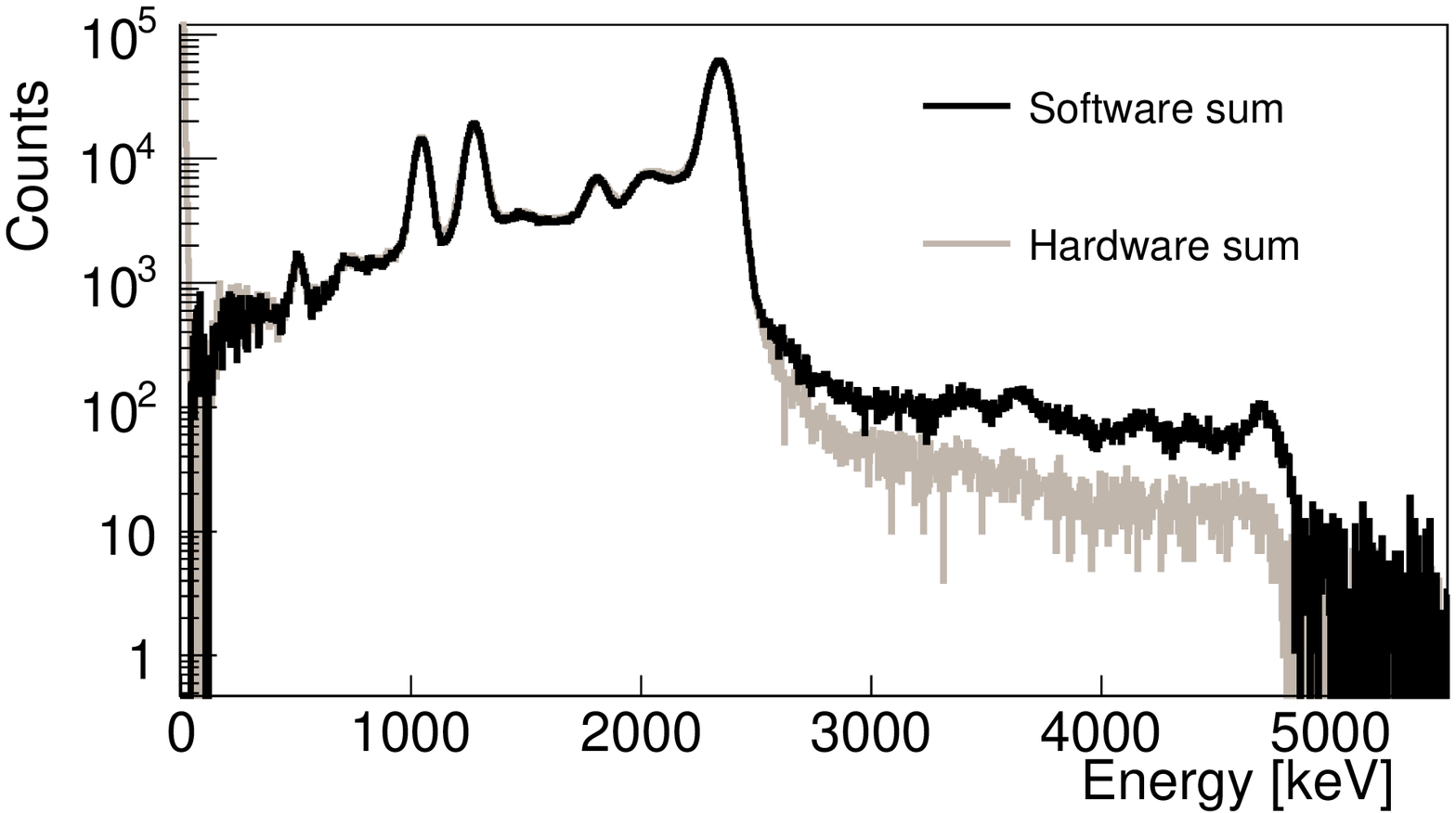}
\caption{Comparison of the hardware sum (grey) and the software sum (black) in DTAS for a $^{60}$Co source (top) and a $^{22}$Na source (bottom).}
\label{hard-soft}
\end{center}
\end{figure}

The software sum reconstructed in this way exhibits the same behaviour as the hardware sum in terms of the non-proportionality of the light yield, as seen in Fig. \ref{hard-soft} for two calibration sources. In both cases the segmented detector behaves as a single crystal detector in terms of the position of the sum peak. The main differences are related to a slightly better resolution in the software sum with respect to the hardware sum due to the gain corrections, and a 
different shape in the pileup region that will be commented on next section. 

\begin{figure}[!h]
\begin{center}
\includegraphics[width=0.5 \textwidth]{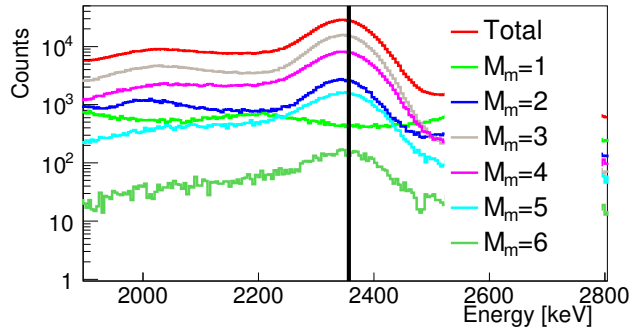} 

\vspace{0.2cm}

\includegraphics[width=0.5 \textwidth]{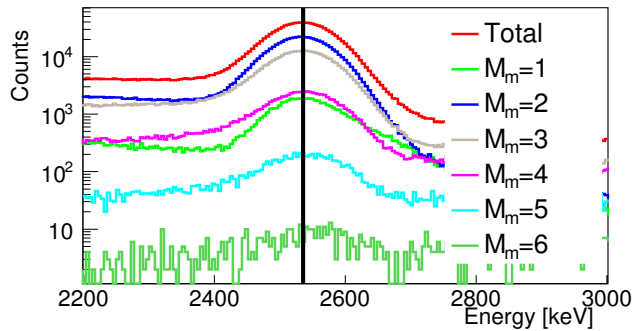} 
\caption{Multiplicity 
spectra
of a $^{22}$Na source (top) and a $^{60}$Co source (bottom). Vertical black lines show the corresponding energy of the sum peak 
for a single crystal detector.
(For interpretation of the references to color in this figure caption, the reader is referred to the web version of this paper.)}
\label{Mult_ok}
\end{center}
\end{figure}

In order to ensure that the treatment of the non-proportionality is correct, we can check the spectra of the multiplicities after this process. In Fig. \ref{Mult_ok}, we show the good alignment of the different $M_m$ multiplicity spectra achieved with this method for a $^{22}$Na source ($M_{\gamma}$=3) and a $^{60}$Co source ($M_{\gamma}$=2), in contrast with the results shown in Fig. \ref{Shift_22Na}. The vertical black lines correspond to the sum peak positions calculated using the shift associated with single crystals according to \cite{TAS_MC} (first row of Table \ref{non-prop_TABLE}): 2296.5~keV+60~keV for the $^{22}$Na source, and 2505.7~keV+30~keV for the case of $^{60}$Co.

We should point out that the procedure followed here solves the misalignment problems
between simulation and experiment encountered in the calibration of the Modular Total Absorption
Spectrometer (MTAS) \cite{non-prop4}, as will be shown in Section \ref{MC_DTAS}.

\section{Summing-pileup calculation}\label{sec-2}

An important source of distortion in the measured spectra is the random superposition of electronic signals within the time length of the ADC gate, due to the statistical nature of the decays. This superposition affects the pulse shape of a single 
detector
leading to so-called pulse-pileup \cite{TAS_pileup}. This 
applies to
individual crystals as well as to the hardware sum of a multi-crystal detector.
In the software sum of a segmented detector the distortion due to the superposition of events within the ADC gate
takes an additional form; namely, the sum of the signals detected in different modules and corresponding to different decays that are, however, stored in the same event. 
Thus to calculate the distortion of the final spectrum
both processes must be taken into account, 
the pulse-pileup (that we will simply call pileup) and the random summing (that will simply be called summing, but has to be distinguished from the traditional use of this term in spectroscopy). We have developed a method to treat the distortion of spectra due to 
summing-pileup
that was already used in previous works \cite{vTAS_PRL,Zak_PRL,vTAS_PRC,Simon_PRC}, and will be detailed here. The 
quality of the  reproduction of this type of spectrum distortion
for a set of calibration sources, listed in Table \ref{sources}, has been studied.

\begin{table}[h]
\begin{center}
\begin{tabular}{c|c}
Source & Rate [kHz] \\ \hline
$^{22}$Na & 4 and 5\\ 
$^{60}$Co & 7 \\
$^{24}$Na & 14  \\ 
$^{137}$Cs & 19  \\
$^{152}$Eu-$^{133}$Ba & 44 \\ 
Background & 3 \\ \hline
\end{tabular}
\caption{Set of sources used in the study of the summing-pileup, and their counting rates in DTAS, together with the environmental background 
rate.
All these measurements were performed with shielding.}
\label{sources}
\end{center}
\end{table}

\subsection{Procedure}

\begin{figure}[!h]
\begin{center}
\includegraphics[width=0.5 \textwidth]{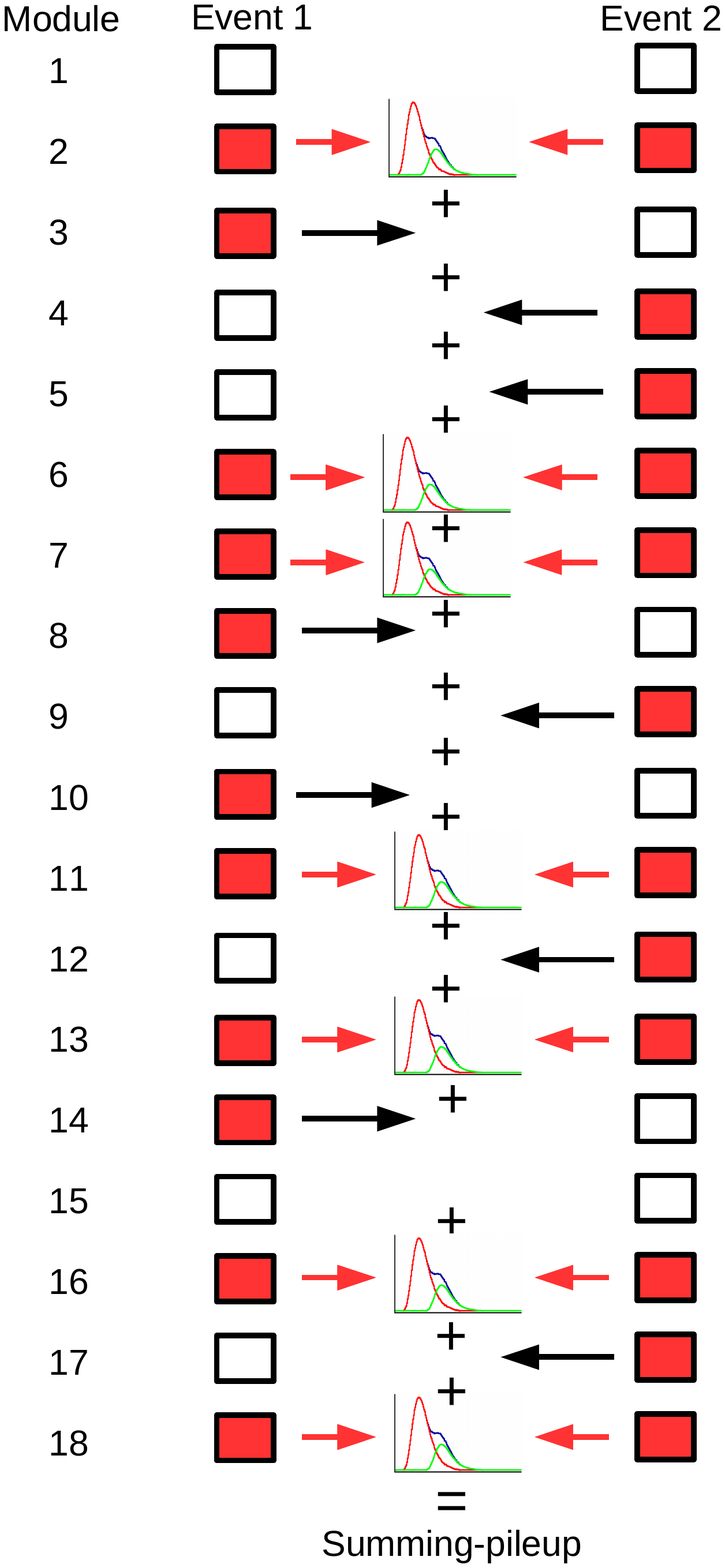}
\caption[First order summing-pileup scheme]{First order summing-pileup scheme. The red squares represent a signal over the threshold stored in the ADC for 
a module in a given event.
When the same module fires in the two events used for the summing-pileup reconstruction, it is processed as a pulse pileup. In any other case, the signals are added giving rise to the summing contribution.}
\label{Summing-pileup_scheme}
\end{center}
\end{figure}

The evaluation of the summing-pileup contamination is based on the event structure of the experimental data, and on the true 
electronic
pulse shape of the individual modules after the MSCF-16 shapers. For the first order 
summing-pileup
calculation, two arbitrary random events are read
from the list-mode event file 
and the time difference between them is sampled randomly within the ADC gate length. If an individual detector has fired in both events, two pulses with their corresponding amplitude are summed, and the maximum within our effective ADC gate $\tau=$ 5.6~$\mu$s (the ADC gate minus the peaking time of the individual signals) is taken, according to \cite{TAS_pileup}. If, on the contrary, the individual detector has only fired in one event, it contributes to the 
summing evaluation. The total summing-pileup is the sum of all contributions, as depicted in Fig. \ref{Summing-pileup_scheme}. 
This procedure assumes implicitly that the distortion
of measured events is small. This approximation is valid if the rate is below 10~kHz. For higher rates
a similar procedure, but based on simulated data, is used as explained in the next subsection.

\begin{figure}[h!]
\begin{center}
\includegraphics[width=0.5 \textwidth]{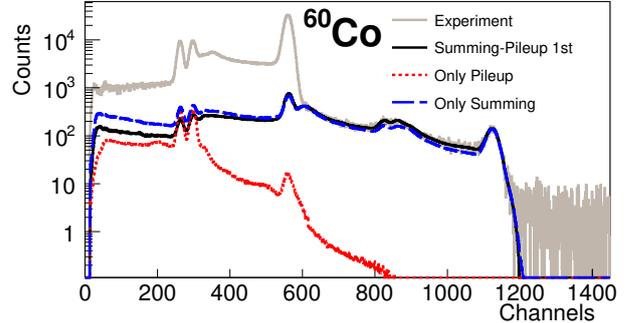}
\caption{First order summing-pileup for the $^{60}$Co source. The contributions of the pulse pileup (dotted red) and the random summing (dashed blue) are separated to show the region where they are affecting the spectrum. (For interpretation of the references to color in this figure caption, the reader is referred to the web version of this paper.)}
\label{Summing-pileup_contributions}
\end{center}
\end{figure}

It is worth mentioning that, in 
our measurements,
the majority of the summing-pileup events 
are coming from the 
summing contribution, as shown in Fig. \ref{Summing-pileup_contributions} for a $^{60}$Co source, where the total summing-pileup contains around 87$\%$ events with only 
summing, $\sim$1$\%$ with only pulse pileup, and $\sim$12$\%$ where both contribute. 

The normalization factors needed to compare 
the calculated summing-pileup contribution
with experimental spectra are obtained from the theoretical expression of Eq. \ref{Pileup_teo_factor}, which is based on the expression 
used
for pileup order $n$ in \cite{TAS_pileup}, but adapted to a segmented detector. Here $\alpha_i$ are the individual counting rates of the 18 crystals and $\tau$ is the length of the effective ADC gate.

\begin{equation}
N^{n}_{theo}=\sum_{i=1}^{18}e^{-\alpha_i \tau}(1-e^{-\alpha_i \tau})^n
\label{Pileup_teo_factor}
\end{equation}

\begin{figure*}[!h]
\begin{center} 
\begin{tabular}{cc}
\includegraphics[width=0.5 \textwidth]{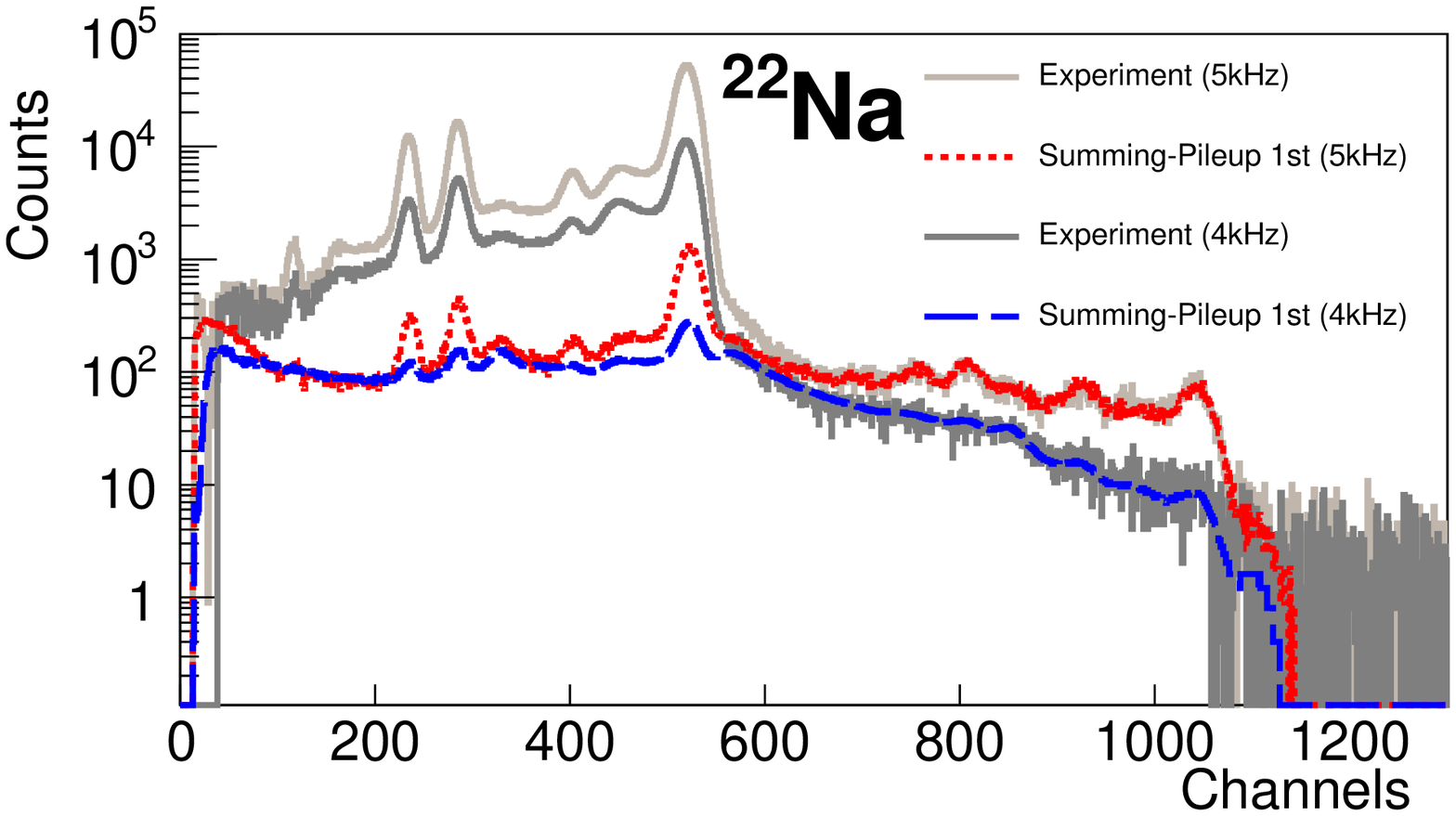}   &
\includegraphics[width=0.5 \textwidth]{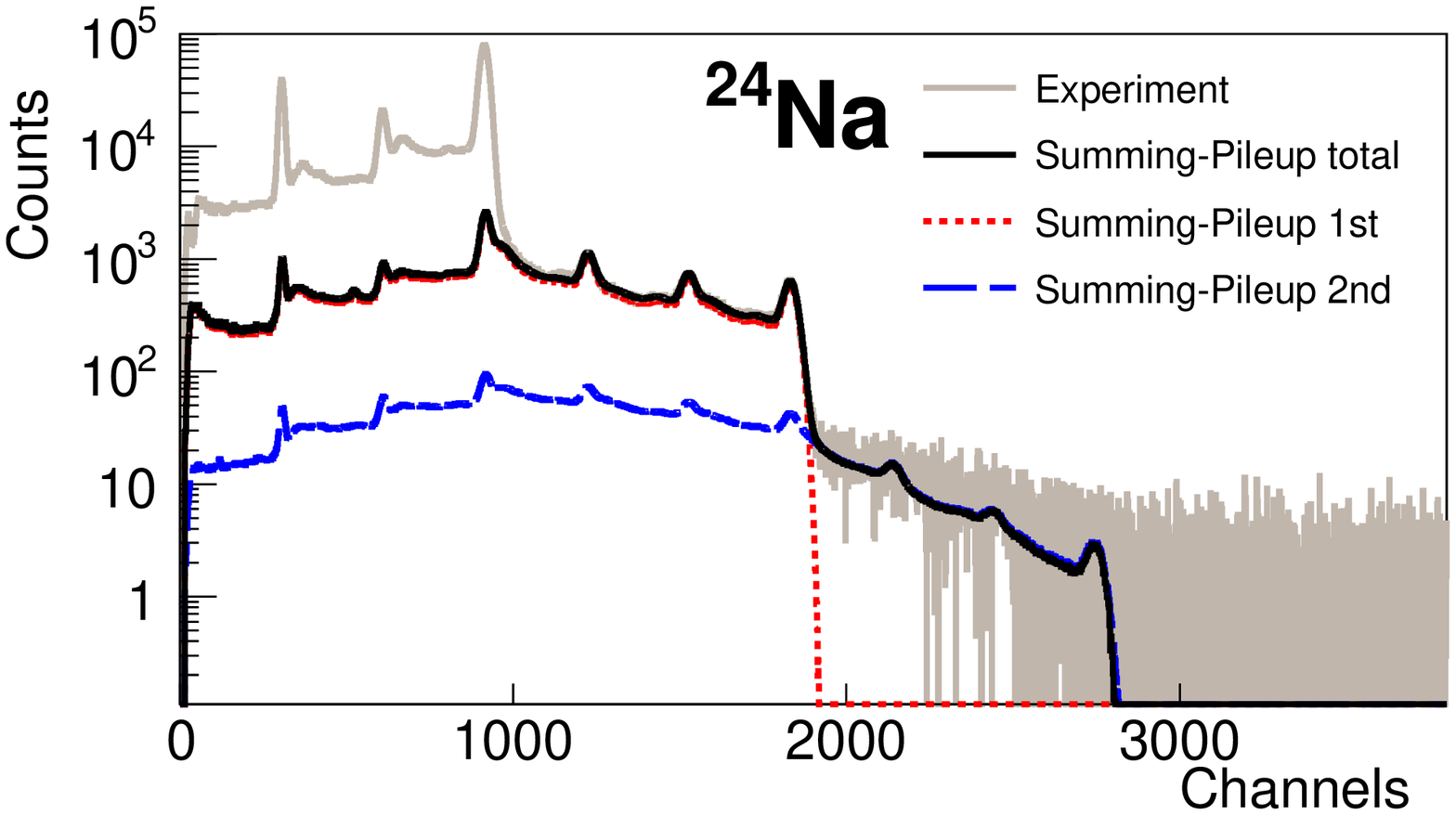} \\
\includegraphics[width=0.5 \textwidth]{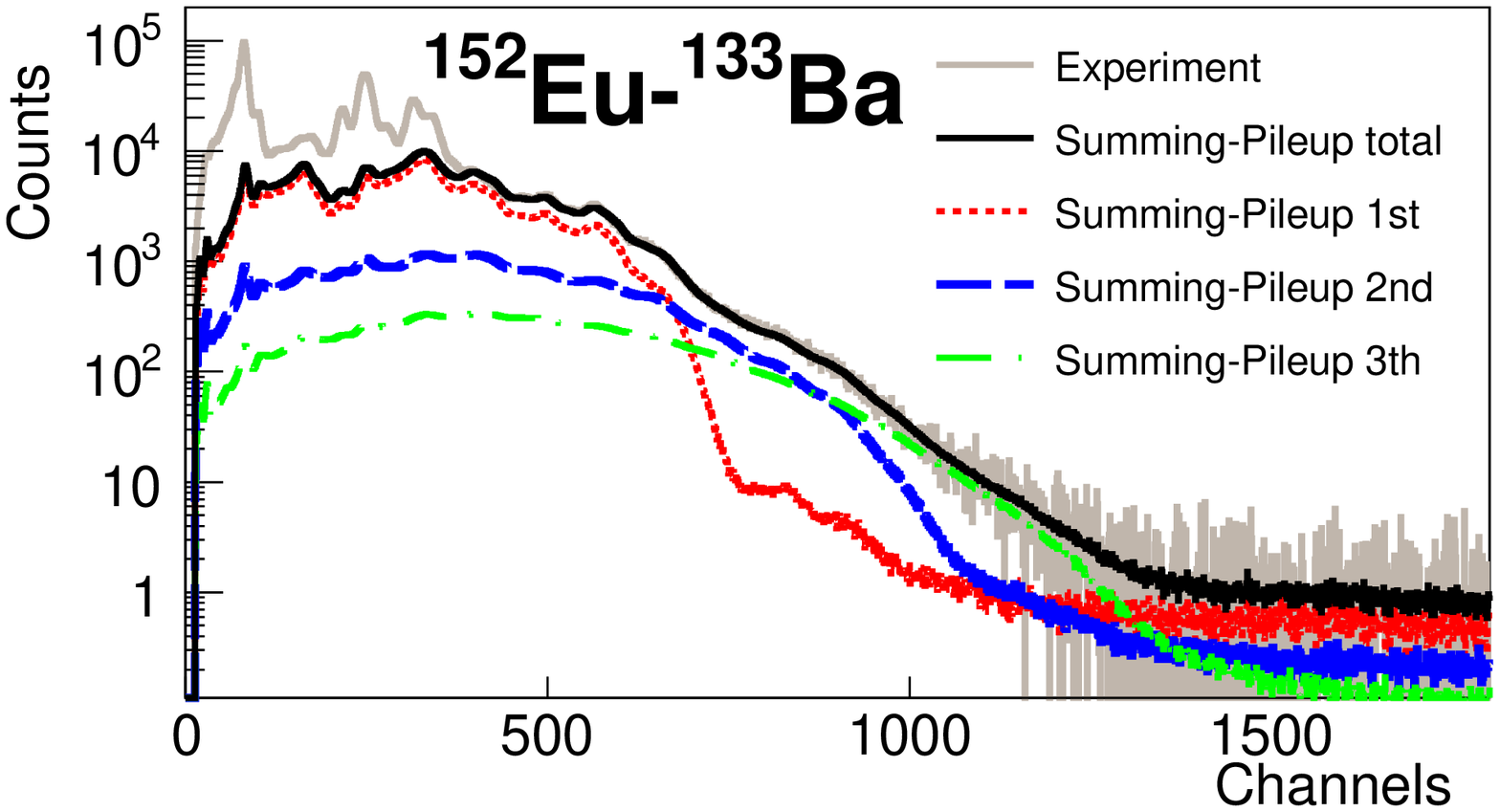} &
\includegraphics[width=0.5 \textwidth]{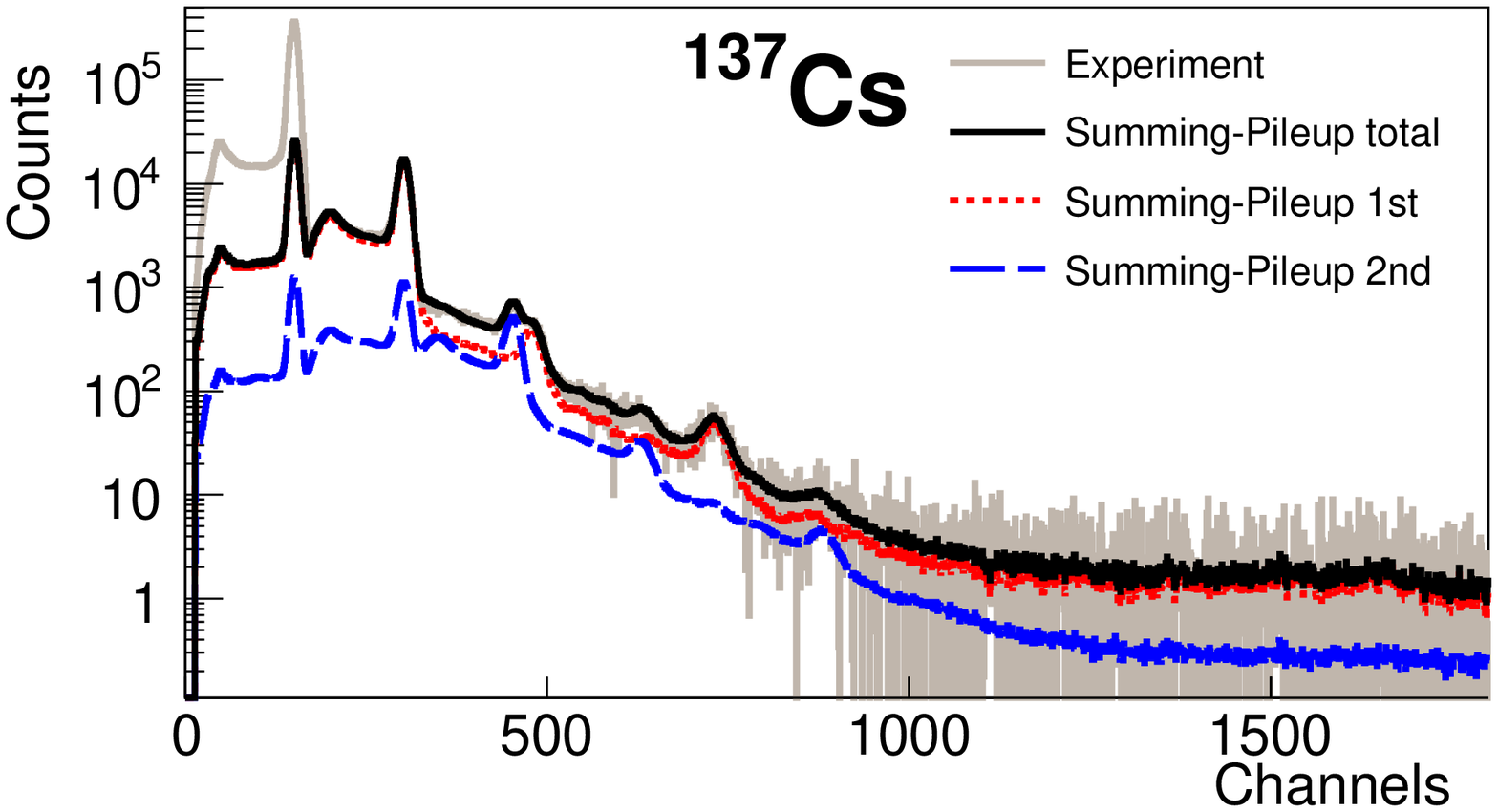}
\end{tabular}
\caption{Summing-pileup contributions for a set of calibration sources. Two sources of $^{22}$Na with different counting rates are compared, normalized by the time of the measurement. For the $^{137}$Cs and the $^{152}$Eu-$^{133}$Ba sources the 
calculated summing-pileup contribution is based on simulated data (see text for details).}
\label{Summing-pileup_sources}
\end{center}
\end{figure*}

When the counting rate is high (approximately above 10~kHz), second order 
summing-pileup
contributions must be evaluated. This is the case for the $^{24}$Na source and the $^{137}$Cs source, whereas for the $^{152}$Eu-$^{133}$Ba source even the third order contribution was needed in order to reproduce the measured spectrum. The procedure in those cases represents just an extension of the method already described. In the second order contribution, for example, three events are taken each time, instead of two. 
The quality of the reproduction of this contamination in the set of calibration sources of Table \ref{sources} can be seen in Figs. \ref{Summing-pileup_contributions} and \ref{Summing-pileup_sources}.

\subsection{Calculation with MC simulated data}

We encountered difficulties
to reproduce the shape of the summing-pileup contribution for 
the
sources with high counting rates, 
$^{137}$Cs and $^{152}$Eu-$^{133}$Ba. 
This is because in these sources, a large fraction of the events that are used for the calculation are already distorted
by summing-pileup. 
However we are unable to distinguish if a measured event is distorted or not.
The way out of this dilemma is to use realistical simulated data.
For this purpose, we simulated the decay of sources with Geant4 \cite{GEANT4} as will be explained in Section \ref{MC_DTAS}
and we stored the relevant information for modules fired in each decay event in a format 
similar to experimental data.
In the simulation the deposited energy is converted into light and the experimental resolution is introduced. The proper light to experimental amplitude calibration is then applied. 

For the calculation of the summing-pileup contribution,
the same procedure explained in the previous section is used with small modifications. 
In particular, 
we have to supplement the simulated data file with a real background data file.
We assume that the summing-pileup distortion in background events is small.
Consequently
only the source-source and source-background summing-pileup contributions are calculated. For this reason the first event is always chosen from the pure source (MC simulation) and the second is taken either from the source, or from the background experimental file. The proportion between source and background for the second event is roughly fixed by the counting rates, and it is a parameter that can be adjusted by looking at the resulting spectra. 

The use of MC 
data
files to reconstruct the summing-pileup contribution has 
proven
to be successful, and the summing-pileup of $^{137}$Cs and $^{152}$Eu-$^{133}$Ba shown in Fig. \ref{Summing-pileup_sources} has been reconstructed by using MC simulated data, 
instead of using the experimental source file, thus validating this method for high counting rates.

\section{Validation of MC simulations}\label{MC_DTAS}

The aim of the TAGS technique is to determine a $\beta$-intensity distribution from an experimental measured spectrum by solving the inverse problem represented by:

\begin{center}
\begin{equation}\label{inverse}
d_i=\sum\limits_{j}R_{ij}f_j + c_i
\end{equation}
\end{center}

\noindent where $d_i$ is the number of counts in channel $i$ of the experimental spectrum, $f_j$ is the number of events that feed level $j$ in the daughter nucleus, and  $R_{ij}$ is the response function of the detector that represents the probability that feeding to the level $j$ gives a count in channel $i$ of the spectrum. 
The sum of all contaminants in channel $i$ is represented by $c_i$.

\begin{figure}[!hbt]
\begin{center} 
\includegraphics[width=0.5 \textwidth]{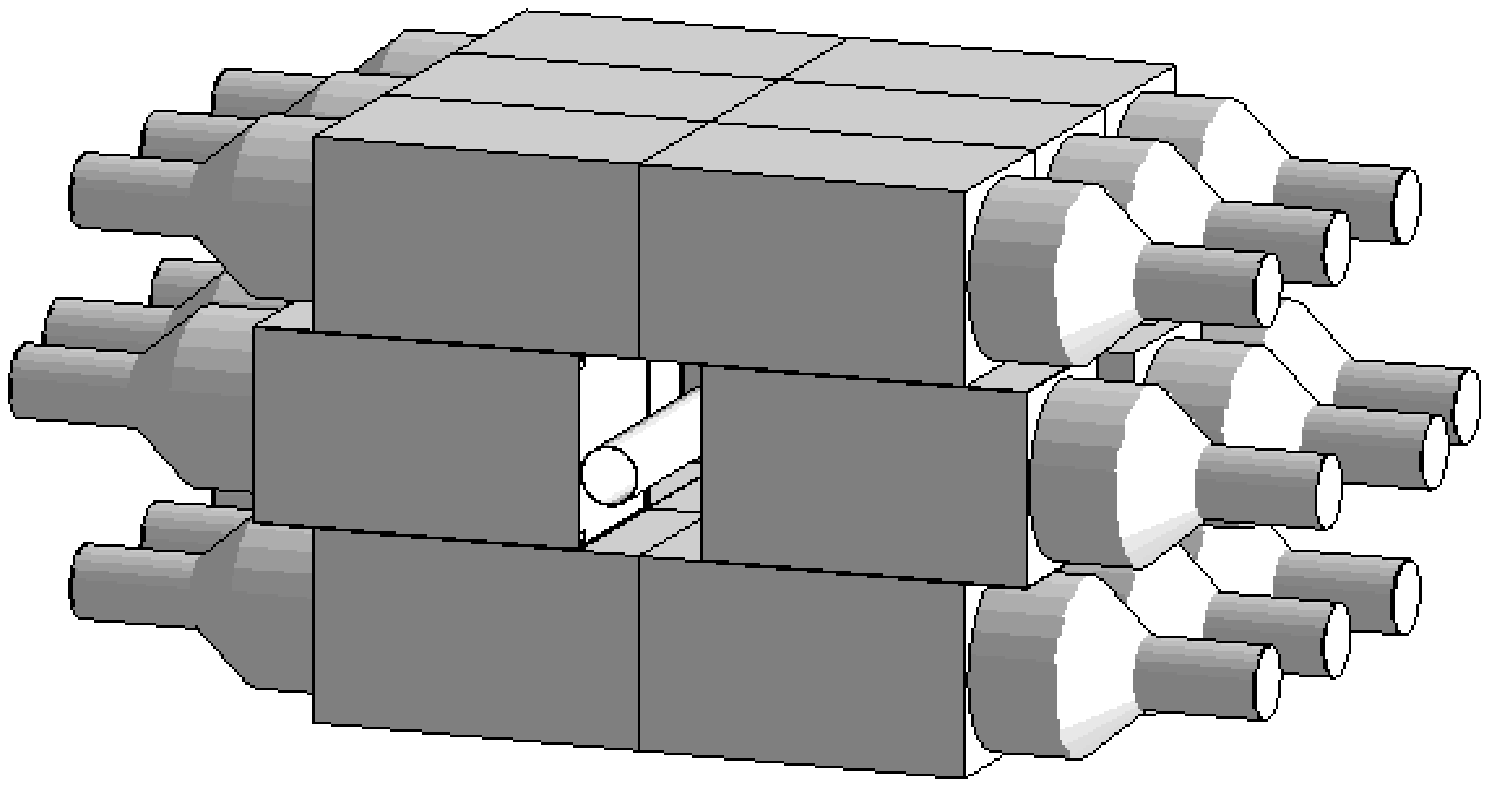} 
\includegraphics[width=0.3 \textwidth]{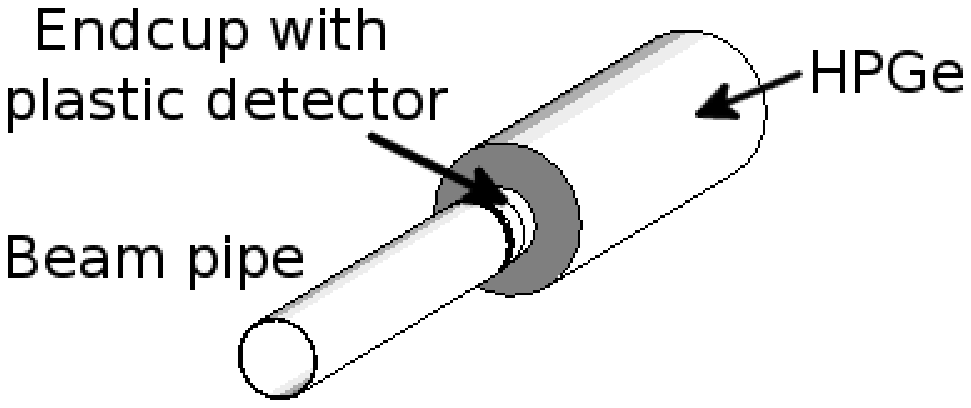}
\caption{Geometry of the set-up implemented in the MC. The DTAS detector in the eighteen-module configuration (top) is represented. Inside we considered the beam pipe, a plastic $\beta$ detector and a HPGe detector (bottom).}
\label{MC_setup_geometry}
\end{center}
\end{figure}

In order to perform this de-convolution 
and obtain
the feeding distribution, a method was developed by the group of Valencia 
\cite{TAS_algorithms} 
which has been successfully applied to a large number of 
cases. An essential ingredient of this process is the determination of the response function, that is unique to each detector and to each decay scheme, and has to be calculated by means of MC codes. For this reason, a mandatory step in the characterization of the detector 
is
to validate the MC
simulation. 
This is achieved by
comparison of simulations with measured calibration sources to verify that the best possible agreement is reached.
The package Geant4 \cite{GEANT4} has been used for this purpose, and the geometry of DTAS has been included in great detail, as shown in Fig. \ref{MC_setup_geometry}. In addition the relevant physics processes involved in particle detection have been incorporated. In particular, the non-proportional light yield in NaI(Tl) has been taken into account according to the parametrization and the procedure detailed in \cite{TAS_MC}. 
In the next subsection we present the results of such a comparison.

\begin{figure}[!hbt]
\begin{center} 
\includegraphics[width=0.5 \textwidth]{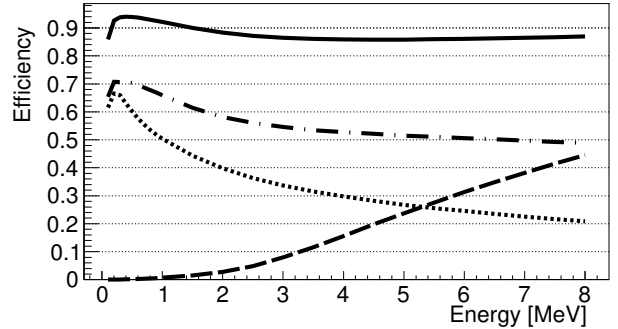} 
\caption{Simulated efficiencies as a function of the energy of the DTAS detector in the eighteen-module configuration. Total $\gamma$ efficiency (solid line) and peak $\gamma$ efficiency (dashed-dotted line) are shown. The peak $\gamma$ efficiency for the array of individual modules is shown as a dotted line. The efficiency for $\beta$ particles as a function of the end-point energy is shown as a dashed line.
Ancillary detectors are included in the simulation (see text for details).}
\label{efficiency}
\end{center}
\end{figure}

The efficiency of the detector for $\gamma$-rays and $\beta$ particles can be 
obtained from the
MC simulations once the geometry and the physics have been validated. In Fig. \ref{efficiency} we 
show the calculated efficiencies 
for the complete set-up used in the commissioning with radioactive beams performed at IGISOL 
with the
eighteen-module configuration \cite{NIMB_DTAS}. The beam pipe, a 3~mm thick plastic scintillator $\beta$ detector with its PMT, and a HPGe detector,
all inserted in DTAS, are included in the geometry.
The efficiency shown is calculated without applying an energy threshold to the individual modules
before reconstructing the sum energy. 
The total efficiency is above 80$\%$ over the whole range, while the peak efficiency at 1~MeV is 66$\%$. When we consider the individual modules in the array, the peak $\gamma$ efficiency at 1~MeV is 50$\%$.



The total and peak efficiencies are limited by the solid angle covered and the amount of both sensitive and dead material.
The solid angle covered can be increased and the dead
material decreased if we remove the HPGe detector and we close the gap between the modules that it ocupies. In this case the efficiencies will increase 
to $\varepsilon_{\gamma}^{T} = 94$\% and $\varepsilon_{\gamma}^{P} = 69$\% respectively at 1~MeV.

In comparison with the Lucrecia spectrometer at ISOLDE \cite{LucreciaTAS} 
and the LBNL spectrometer at GSI \cite{NIMB_TAS_GSI}, both single crystal spectrometers, 
the efficiency of the present eighteen module configuration is similar.
When compared with the recently built MTAS spectrometer \cite{MTAS_effic} our peak efficiency is 9\% smaller
at 1~MeV and 25\% smaller at 3~MeV. 
This difference is a consequence of the much larger NaI(Tl) volume in MTAS, which is a factor of
2.5 larger than DTAS. It should be noted that the smaller efficiency of DTAS does not affect its performance
as a total absorption spectrometer. A nice example is provided by the decay of $^{137}$I which has been
measured by DTAS \cite{NIMB_DTAS} (see also subsection 4.2) and MTAS \cite{MTAS_137I}. Figure 4
in \cite{MTAS_137I} compares the spectrum measured with the full MTAS with the spectrum measured
with the sub-detector consisting of the 7 most central modules. This sub-detector is equivalent to
DTAS in volume (about 100 litres) and efficiency. As can be observed the differences are minimal
except for the contamination induced by the interaction of delayed neutrons emitted in the decay, which
is much larger in the full MTAS. The reason why the much larger volume brings a seemingly small
effect is to be found in the complex de-excitation pattern, with relatively large cascade multiplicities,
and the effect of $\beta$ penetration which tends to wash out the increase in single $\gamma$-ray
peak efficiencies. Such a consideration was taken into account during the design of DTAS
when choosing the detector size \cite{DTAS_design}.

\subsection{Reproduction of the calibration sources}

\begin{figure*}[h]
\begin{center} 
\begin{tabular}{cc}
\includegraphics[width=0.5 \textwidth]{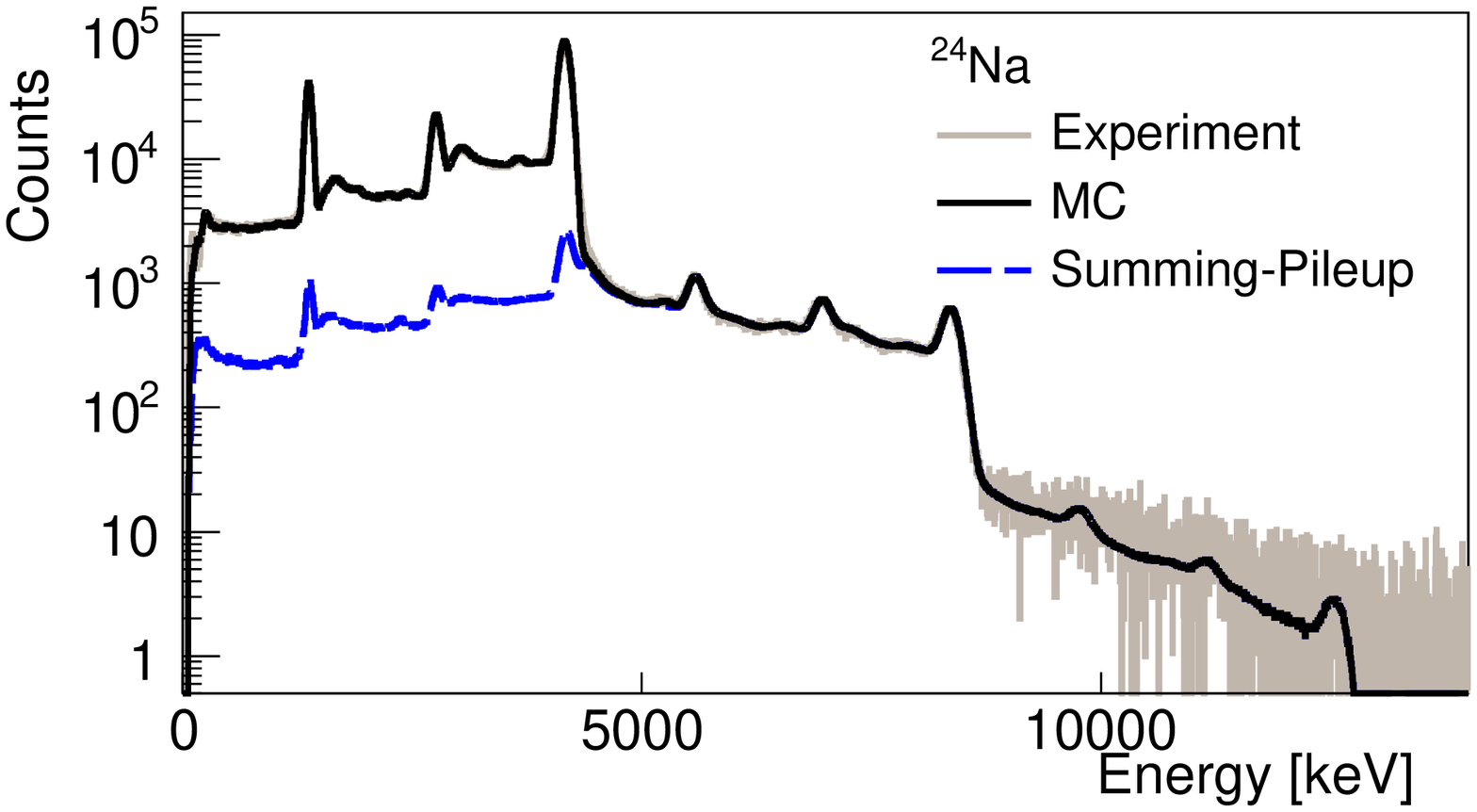} &
\includegraphics[width=0.5 \textwidth]{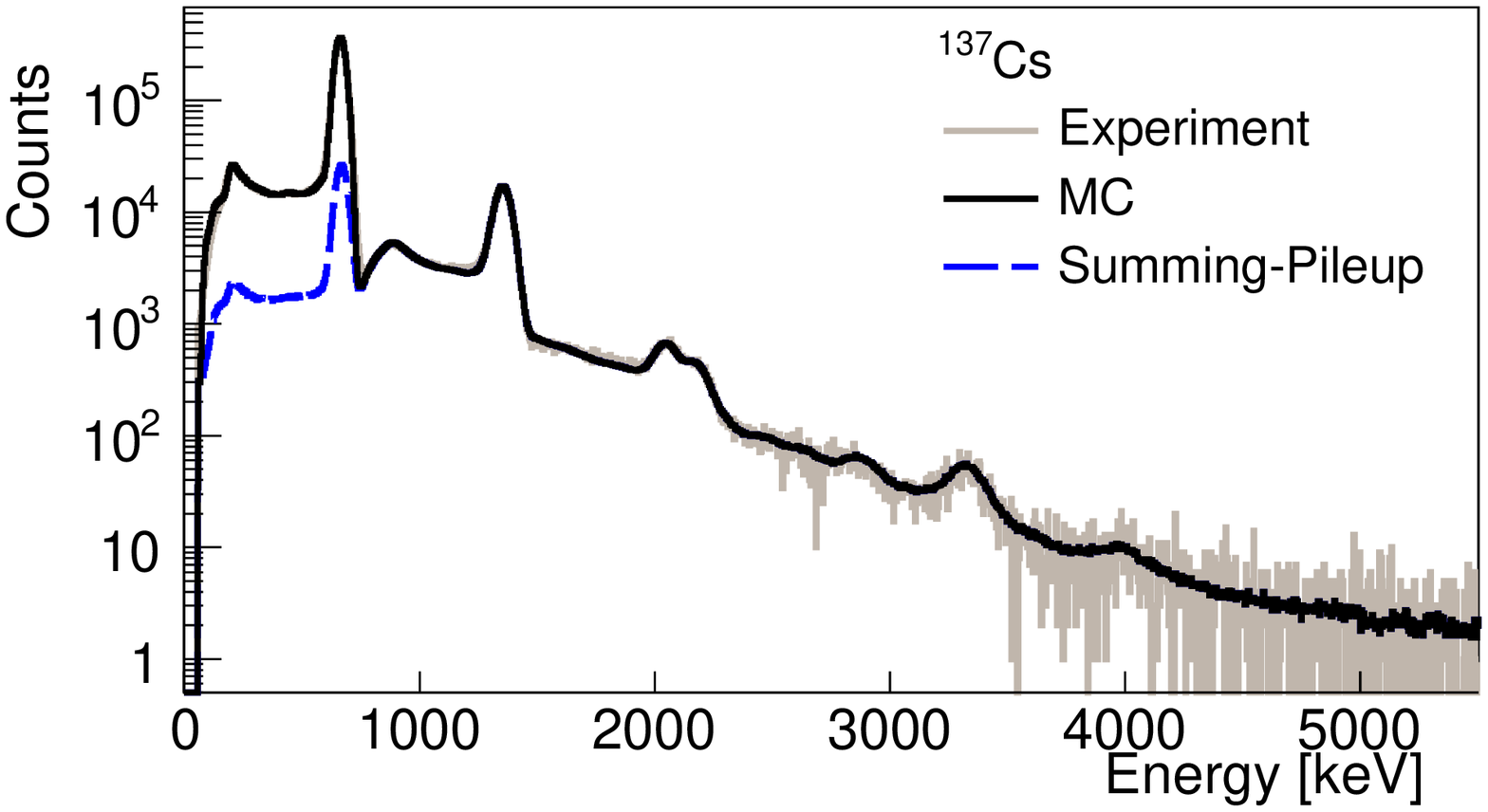} \\
\includegraphics[width=0.5 \textwidth]{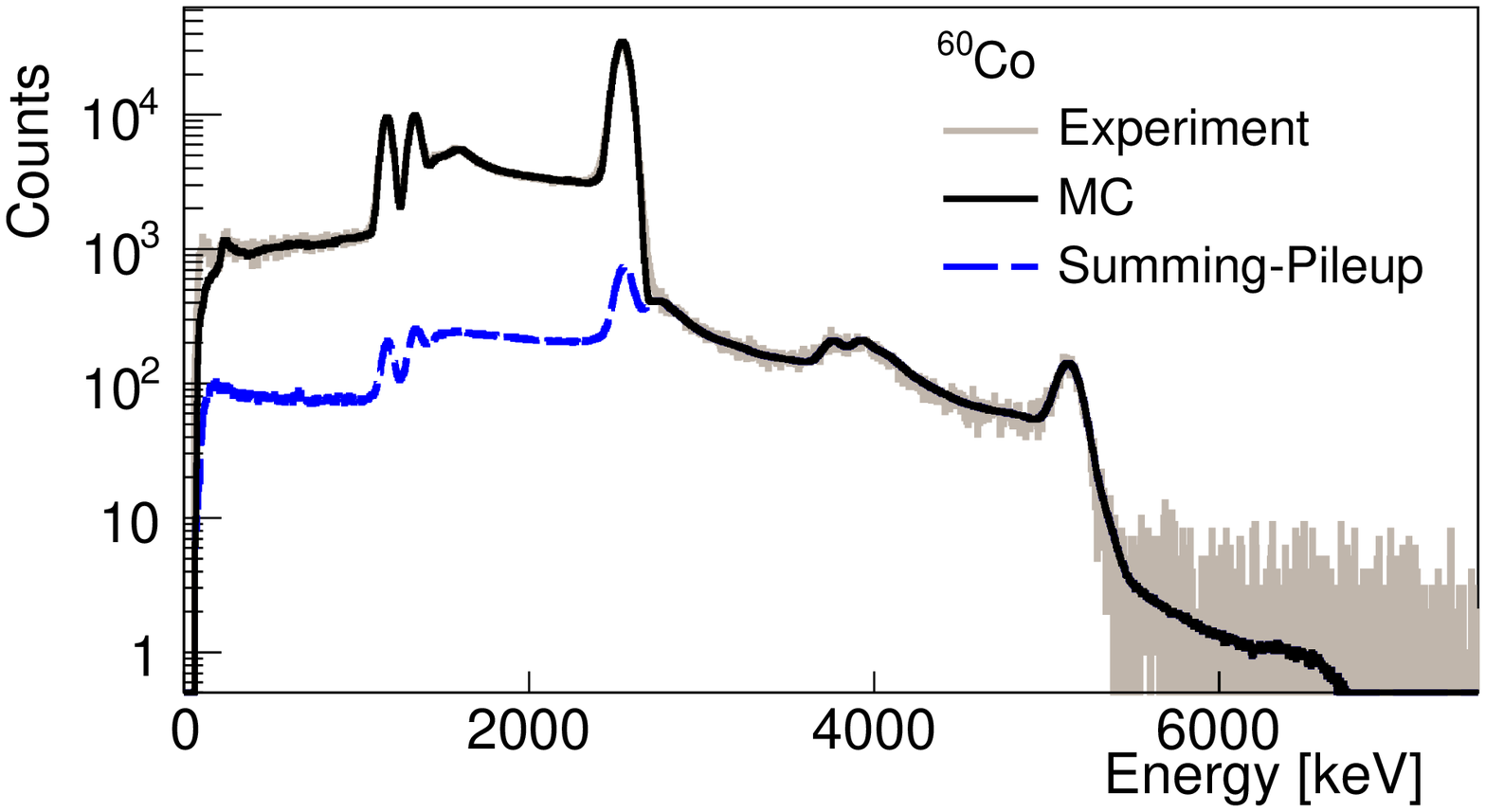} &
\includegraphics[width=0.5 \textwidth]{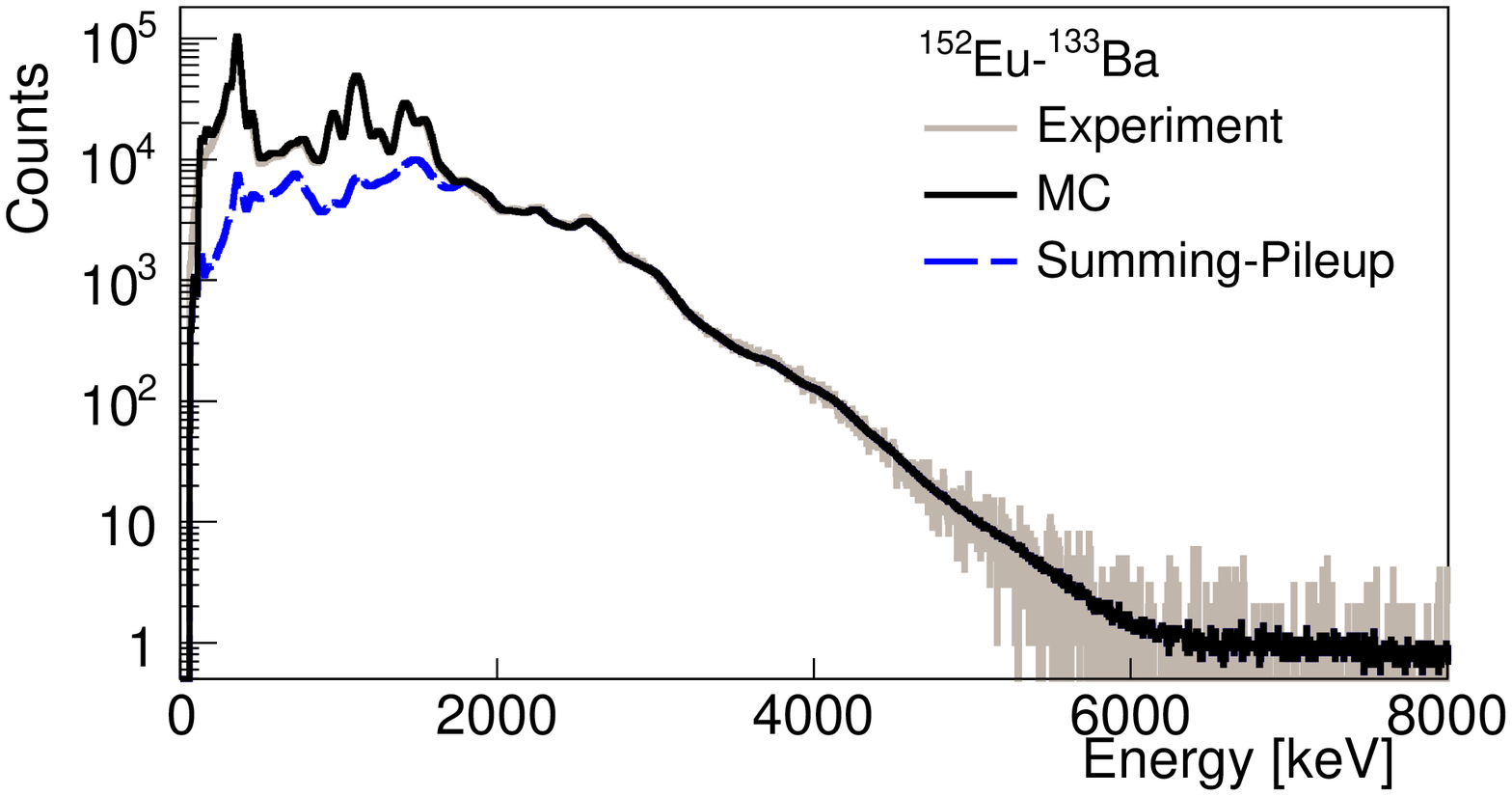}
\end{tabular}
\caption{Experimental 
spectra
of calibration sources after subtracting the environmental background (solid grey) compared with the MC simulations (solid black) taking into account the summing-pileup contamination (dashed blue).}
\label{MC-exp}
\end{center}
\end{figure*}

In this subsection we compare the results of the MC simulations with measurements for the calibration
sources in Table \ref{sources}. In the comparison, the different sources of contamination in the 
measured spectra are taken into account. The environmental background is subtracted from
the measured spectra. However we choose to show explicitly the summing-pileup contribution, calculated
as described in the previous section, adding it to the MC simulated spectra for the comparison.
We use the DECAYGEN event generator \cite{TAS_decaygen} to generate the primary particles in the MC simulations. As can be observed in Fig. \ref{MC-exp}, we obtain an excellent reproduction of the experimental spectra.

One of the key features of DTAS is its segmentation. This allows one to obtain much richer information, provided by the
energy spectra of the individual modules and more importantly by the sum energy spectra gated with different conditions 
on the number of modules that fired ($M_m$). These additional spectra are sensitive to the details of the de-excitation
cascades (energies and multiplicities $M_{\gamma}$). 

\begin{figure*}[h]
\begin{center} 
\begin{tabular}{cc}
\includegraphics[width=0.5 \textwidth]{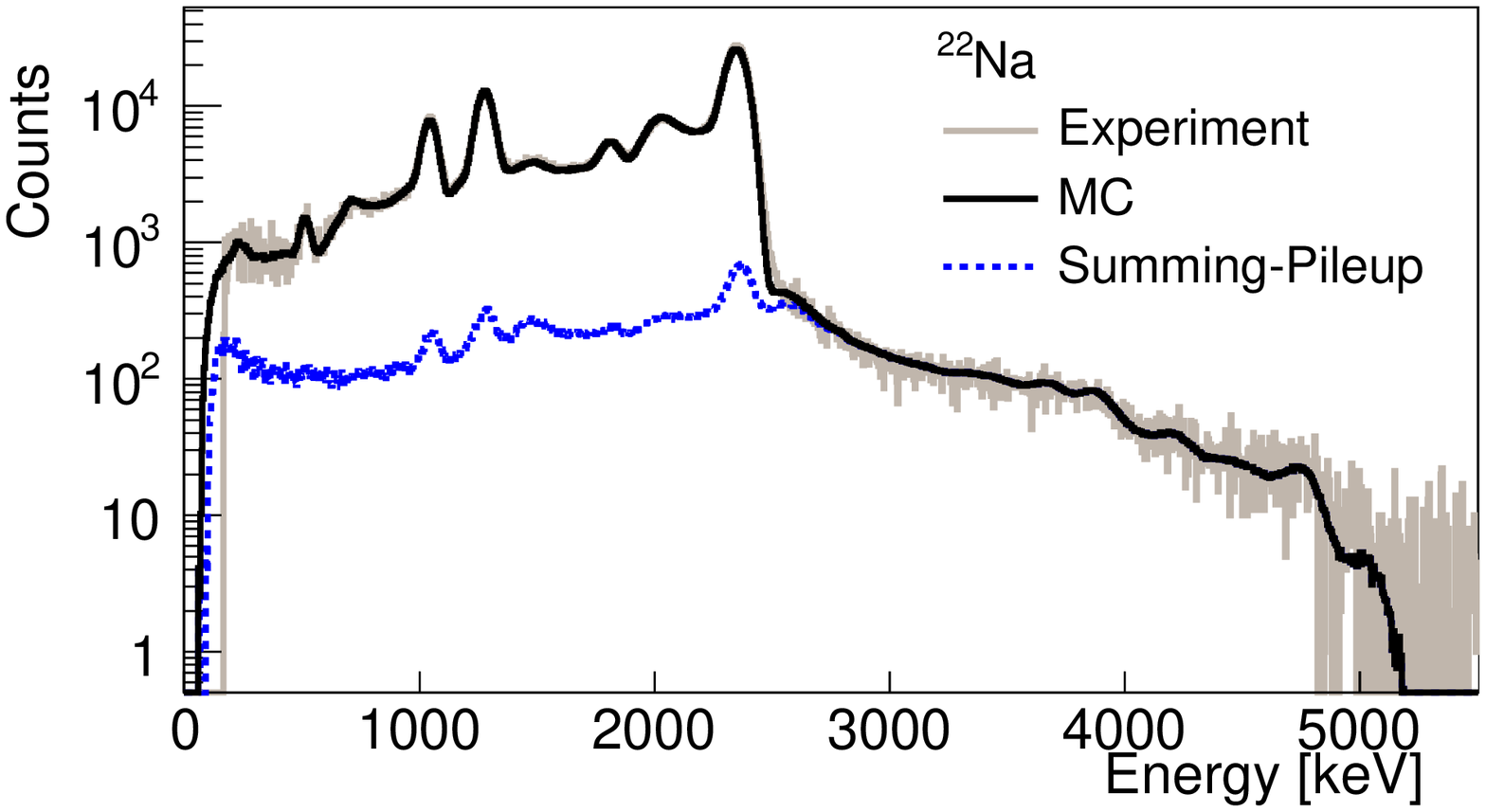} &
\includegraphics[width=0.5 \textwidth]{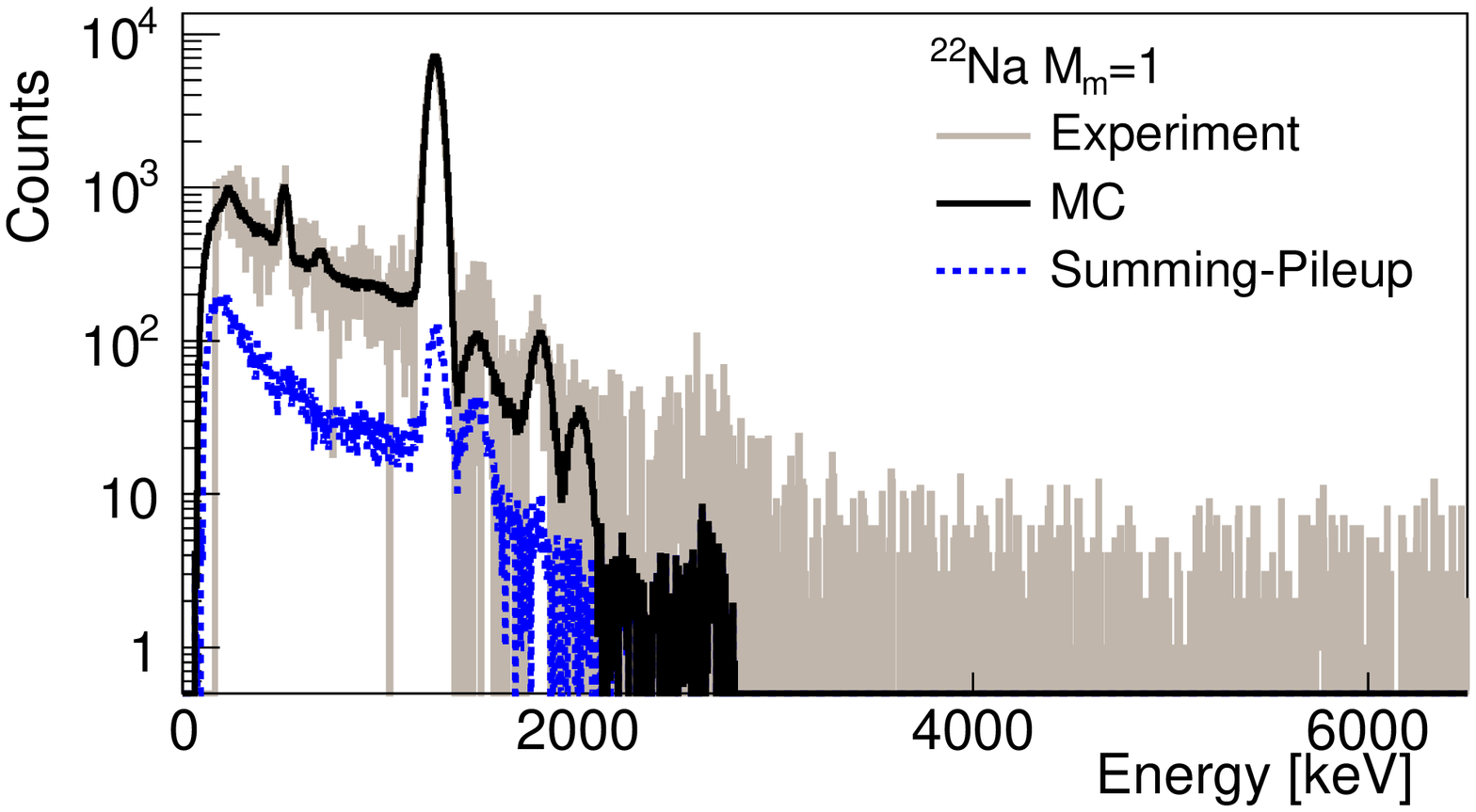} \\
\includegraphics[width=0.5 \textwidth]{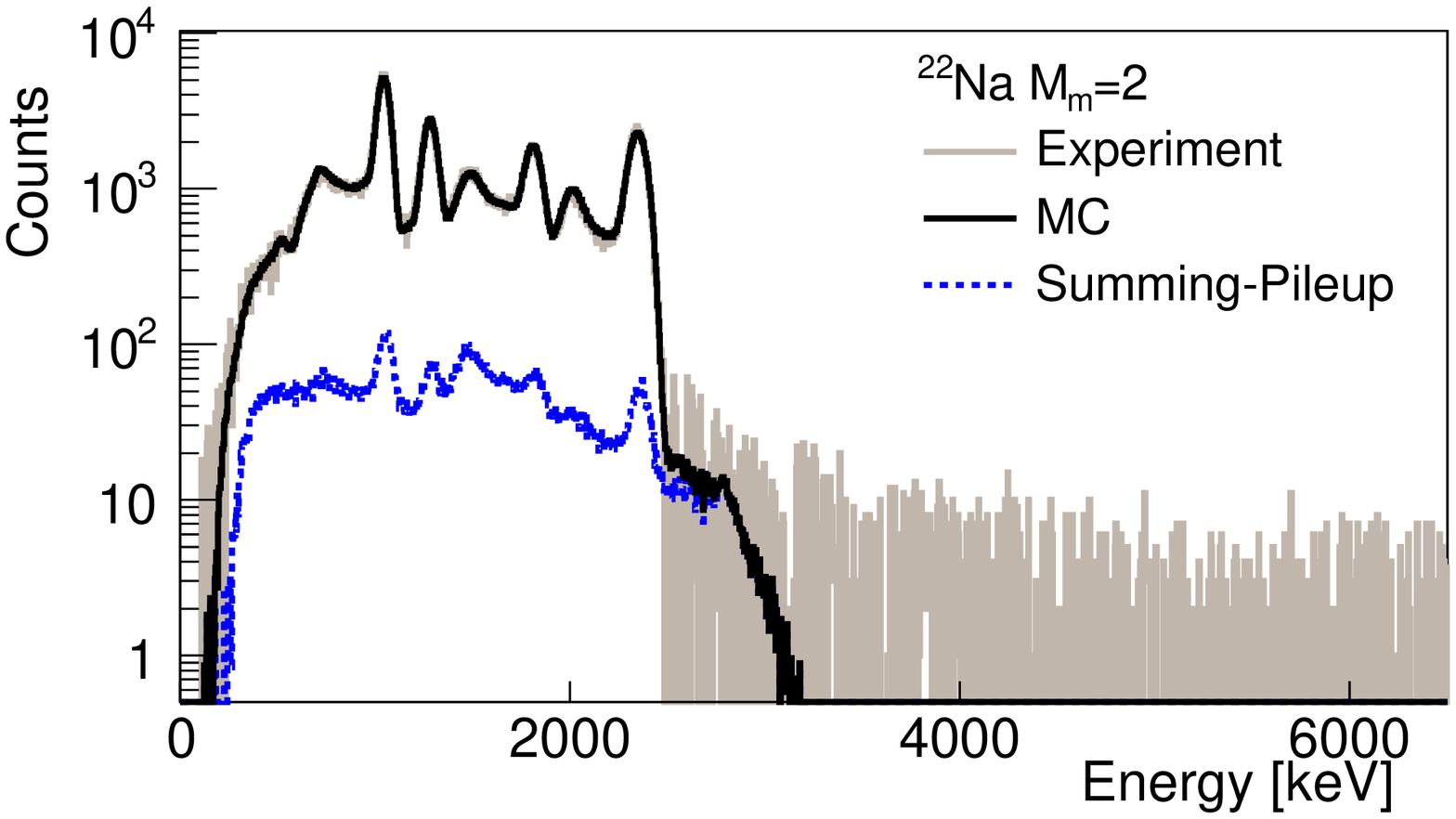} &
\includegraphics[width=0.5 \textwidth]{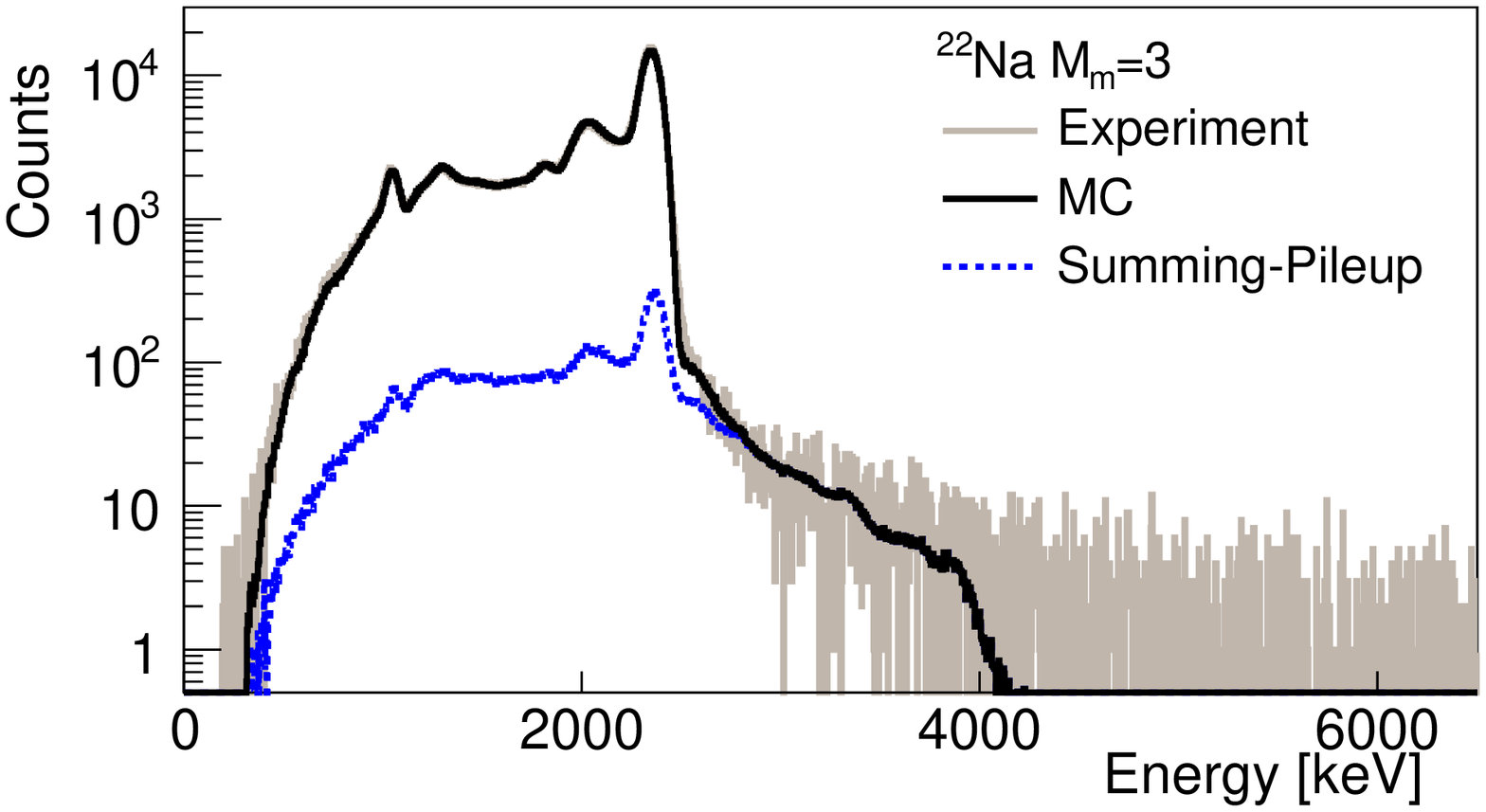} \\ 
\includegraphics[width=0.5 \textwidth]{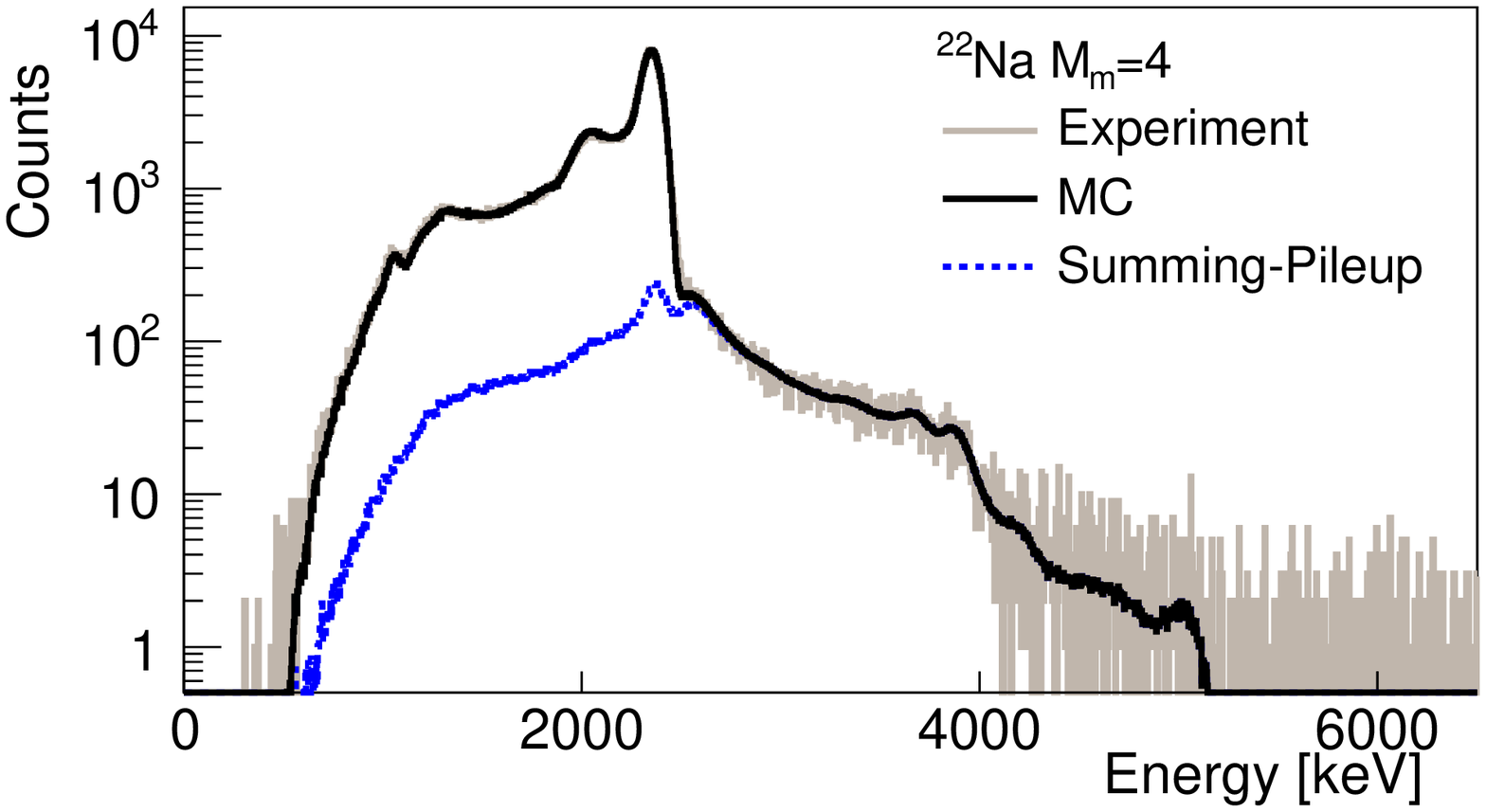} &
\includegraphics[width=0.5 \textwidth]{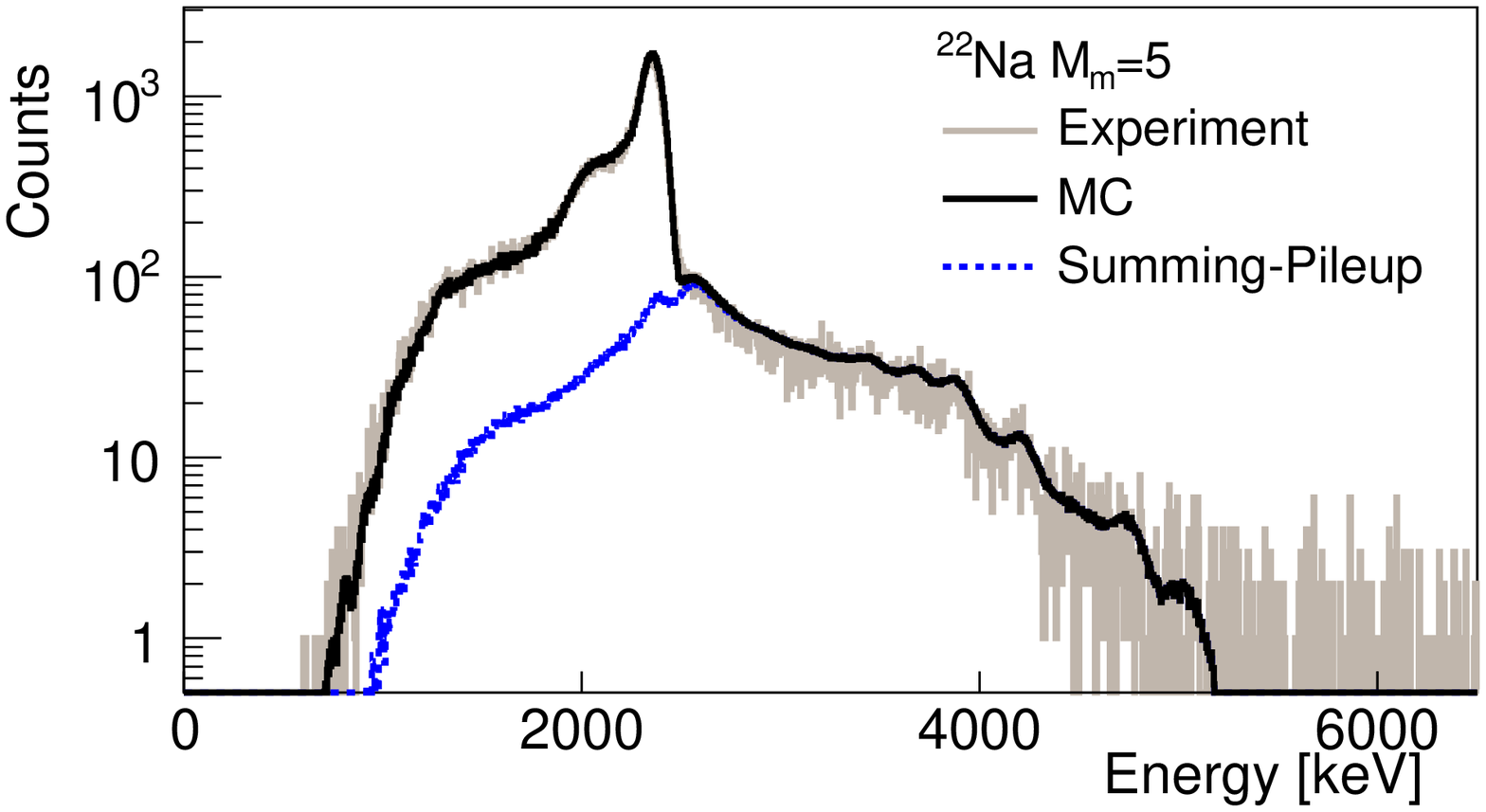} 
\end{tabular}
\caption{$^{22}$Na experimental 
spectra
after subtracting the environmental background (solid grey) compared with the MC simulations (solid black) taking into account the summing-pileup contamination (dashed blue). 
The sum energy spectrum without conditions and with a condition on 
module multiplicity $M_m$ from 1 to 5 is shown.}
\label{MC_multiplicities}
\end{center}
\end{figure*}

In the case of laboratory sources with known decay schemes the multiplicity  information provides a more stringent 
test of the accuracy of the MC simulation, both of the geometry and the physical processes included.
As can be seen in Fig. \ref{MC_multiplicities} for the $^{22}$Na source an excellent agreement is obtained
proving that we have the MC simulations under good control.
It should be noted that all calculated spectra shown in Fig. \ref{MC_multiplicities} are obtained simultaneously
using the same energy calibration, a common normalization factor and the same summing-pileup calculation.

\subsection{Reproduction of neutron interactions}

The emission of $\beta$-delayed neutrons in the decay of exotic neutron-rich nuclei is a source of background for total absorption spectrometers like DTAS.
Neutrons
interact with 
detector materials
producing $\gamma$-rays, either in an inelastic reaction or after neutron capture. 
When detected these $\gamma$-rays
are indistinguishable from  $\beta$ delayed $\gamma$-rays. 
The ability to reproduce this type of background correctly with MC simulations is fundamental for the analysis of 
TAGS spectra from $\beta$-delayed neutron emitters. The issues related to the simulation of neutron interactions 
in inorganic scintillators, and in particular the use of the Geant4 simulation tool,  have been discussed before
\cite{neutrons,DTAS_design}. One important item is the quality of the information in nuclear data bases concerning
reaction cross sections for all the materials encountered by the neutrons. In the present simulations we used the library ENDF-VII.0, which gave good results before. 
The data from this library were converted into the G4NDL data format \cite{CIEMAT_neutrons}. 
Another important item is the description of the $\gamma$-ray de-excitations
of excited states resulting from neutron interactions. We replace the standard capture cascade generator of Geant4,
which is rather schematic, with a generator that uses the statistical model to describe realistically the multiplicity 
and energy distribution of the cascades. In the case of inelastic scattering we use the standard Geant4 
PhotonEvaporation data base which relies on evaluated spectroscopic data \cite{ENSDF}.



We have studied the decay of two $\beta$-delayed neutrons emitters measured at IGISOL:
$^{137}$I and $^{95}$Rb. Preliminary results for $^{137}$I were presented in \cite{NIMB_DTAS}. Relevant decay information for these nuclei is well established in the data bases. The values from ENSDF \cite{ENSDF} of $Q_{\beta}$, neutron separation energy in the daughter $S_n$  and neutron emission probability $P_n$ are given in Table~\ref{bdn}. We recently performed an accurate $P_n$ measurement for both isotopes using the BELEN neutron counter \cite{NIM_BELEN} which gave 9.08(14)$\%$ for $^{95}$Rb and 7.76(14)$\%$ for $^{137}$I, close to the values in the table.

\begin{table}[h]
\begin{center}
\begin{tabular}{cccc}
Isotope &  $Q_{\beta}$ [MeV] & $S_n$ [MeV] & $P_n$ [\%]\\ \hline
$^{137}$I & 6.027(9) & 4.02556(10) & 7.14(23) \\ 
$^{95}$Rb & 9.228(21) & 4.348(7) & 8.7(3) \\ 
 \hline
\end{tabular}
\caption{Properties of $\beta$-delayed neutron emitters used to test MC simulations.}
\label{bdn}
\end{center}
\end{table}


We compare with the simulation the sum energy spectra gated with $\beta$ particles detected in
a thin plastic scintillator. These spectra are free from environmental background and are affected 
by the end-point energy dependence of the $\beta$ efficiency which suppresses the $\beta$ decays
to states close to $Q_{\beta}$. In order to take this effect properly into account
an event generator was implemented \cite{vTAS_PRC} that reproduces the known sequence of 
$\beta$-neutron-$\gamma$ emission in the decay. It reproduces also the measured
neutron spectra obtained from the ENDF/B VII.1 database, based on the work in \cite{Brady_thesis}. 
The generator requires the reconstruction of the $\beta$ intensity distribution from the measured neutron spectra using the information on neutron branchings to the excited levels in the final nucleus, $I_n$. The generator uses the associated $\gamma$ branchings, $I_{\gamma}$, as well. Both $I_n$ and $I_{\gamma}$ data were retrieved from the ENSDF database \cite{ENSDF}. 

\begin{figure}[h]
\begin{center} 
\includegraphics[width=0.5 \textwidth]{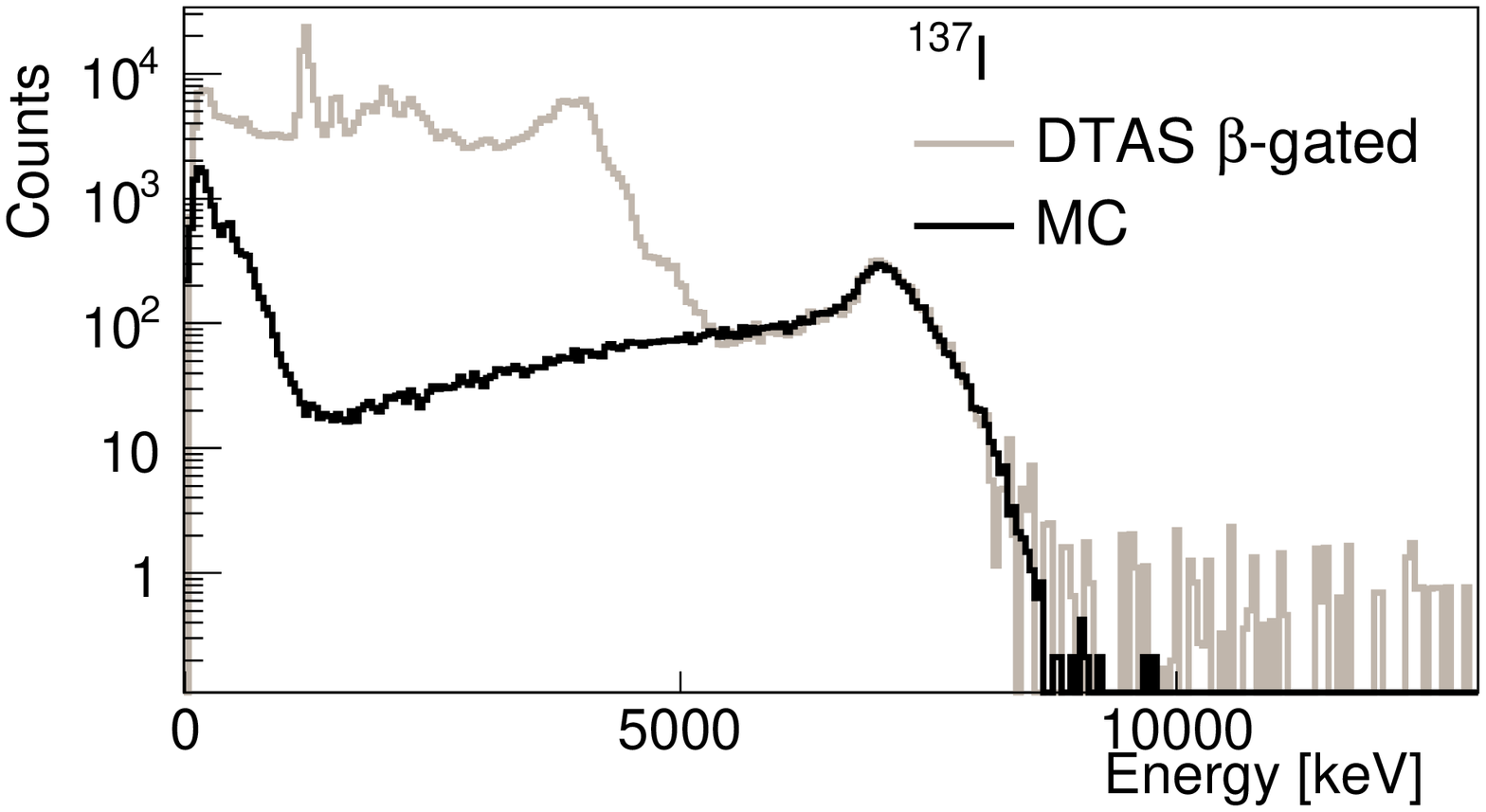} 
\includegraphics[width=0.5 \textwidth]{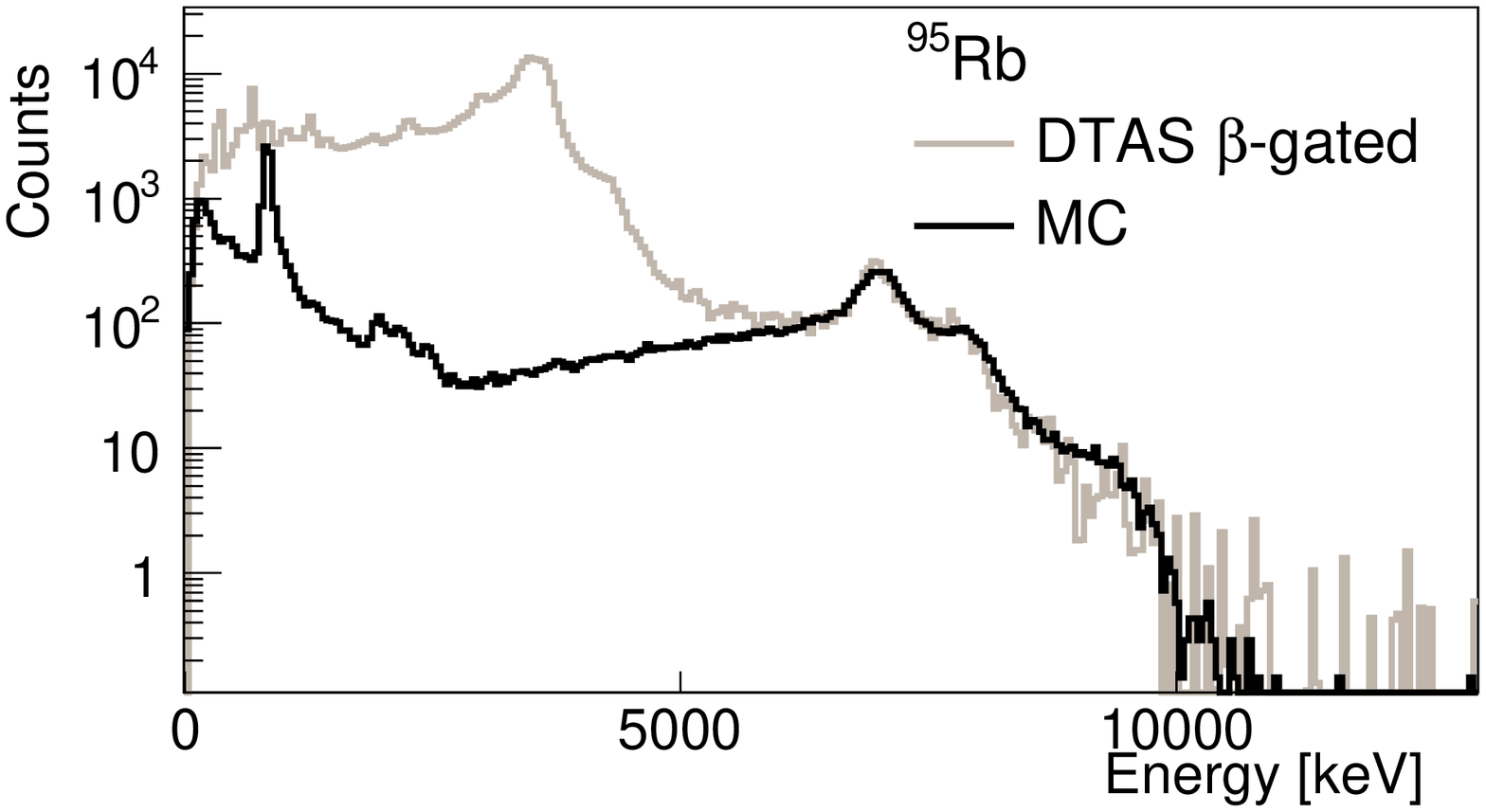}
\caption{
Simulation of the $\beta$-delayed neutron decay branch for 
neutron emitters measured in the commissioning of DTAS at IGISOL: $^{137}$I (top) and $^{95}$Rb (bottom). Experimental $\beta$-gated spectra (in grey) 
are compared to simulations (in black).}
\label{neutrons_MC}
\end{center}
\end{figure}

In the simulations a time window for accumulation of the energy deposited after multiple neutron interactions was applied. This window takes into account the existence of a delay between neutron-induced $\gamma$-rays 
and the prompt $\beta$ signal. A window of 500~ns was employed in accordance with the experimental coincidence time window between DTAS and the $\beta$ plastic detector. This window ensures the collection of all the energy deposited. Figure \ref{neutrons_MC} shows the comparison of measured and simulated spectra for the two $\beta$-delayed neutron emitters studied. As can be observed
the reproduction of the gross structure above 6.8 MeV, mainly due to neutron capture in the
iodine ($^{127}$I) in the crystal, is very good. In the case of $^{137}$I the shape
of the structure depends on the $\beta$-delayed neutron energy spectrum.
It should be noted however that 
in the case of $^{95}$Rb this structure
includes partial summing of capture $\gamma$-rays with
$\gamma$-rays emitted from excited states populated
after neutron emission.
The strongest of these $\gamma$-rays is also visible as a peak in the simulated 
spectra at 837~keV superimposed on the $\gamma$-ray background from neutron inelastic collisions.
We found that the shape of the gross structure is quite sensitive to the de-excitation pattern after
neutron emission. For $^{95}$Rb, the use of the evaluated  decay scheme available  in ENSDF
produced a wrong shape for the spectrum. However when we use the de-excitation scheme
in $^{94}$Sr measured by Kratz et al. \cite{Kratz_95Rb} a good reproduction was obtained
as can be seen in Fig. \ref{neutrons_MC}. Thus except when the neutron emission
proceeds entirely to the ground state, the effect of the neutron energy distribution on the shape of the capture peak is obscured by the final nucleus $\gamma$ spectra. Since the latter is often unknown or poorly known, it seems difficult to obtain reliable information about the shape of the $\beta$-delayed neutron spectrum from TAGS spectra in the general case.

Ideally the normalization of the $\beta$-delayed neutron contribution to the total spectrum should be determined by the $P_n$ value. From the normalization of the simulated and measured counts in the capture bump at 6.8~MeV we have obtained $P_n$ values of 6.8$\%$ and 10.9$\%$ for $^{137}$I and $^{95}$Rb respectively, after taking into account all the contaminants (summing-pileup and activity of the descendants). Note that in comparison with the numbers given in Table \ref{bdn}, we found a 5$\%$ smaller value for $^{137}$I, while for $^{95}$Rb a 25$\%$ larger value is obtained. Compared to our recently measured values \cite{NIM_BELEN} the differences are -12$\%$ and +20$\%$ respectively. We studied the dependence of the extracted $P_n$ on the length of the time window applied to the experiment (coincidence gate) and to the MC simulation. We found that in the range 300-500~ns the results were stable within 3$\%$. It should be noted that the value we obtain is 14$\%$ lower than the value of 7.9(4)$\%$ obtained by a similar procedure with MTAS \cite{MTAS_137I}. In view of this discrepancy and the fact that for $^{95}$Rb we also obtain a large difference but of opposite sign, we conclude that further investigations are needed before deciding on the reliability of $P_n$ extraction from TAGS spectra \cite{MTAS_137I}. 

In any case, the key point for us is the reproducibility of the shape of the spectra of the $\beta$-delayed neutron contamination that affects the extraction of $I_{\beta}(E_x$) from the analysis of TAGS spectra. A proper determination of this background component is particularly relevant when extracting an accurate value for the $\beta$ intensity above the neutron separation energy that proceeds by $\gamma$ emission, $P_{\gamma}$ \cite{vTAS_PRC}. The investigation of $\gamma$/neutron competition from neutron unbound states is a topic of current active research. The importance of the correction of the background due to $\beta$-delayed neutrons for the determination of $P_{\gamma}$ using TAGS spectrometers made of NaI(Tl) can not be overlooked. This material has a large capture cross-section resulting in large $\beta$-delayed neutron detection efficiencies, of the order of 40$\%$. We note, in particular, that this correction has been ignored in a recent measurement of $P_{\gamma}$ for $^{70}$Co decay with the SuN spectrometer \cite{PRL_70Co} and might change their result significantly.

The sensitivity of the MC simulation of the $\beta$-n decay contamination to the knowledge of the decay (neutron spectrum and $\gamma$-cascades 
after neutron emission) represents a challenge for very neutron-rich nuclei in the general case where this information 
is poorly known or not known at all.  Given that $\gamma$-rays produced by neutron interactions are delayed  
with respect to $\beta$-particle emission one can use timing information to discriminate between these signals \cite{DTAS_design}. We have tested this idea for $^{137}$I and $^{95}$Rb with reasonable results, as will be shown in a forthcoming publication. However, this type of time discrimination cannot be applied for the $\gamma$-ray de-excitation in the final nucleus after neutron emission since they are prompt with respect to the $\beta$-particles. The best option here seems to use the spectrometer itself to obtain information about this type of contamination as was suggested in \cite{MTAS_effic}. The modularity of DTAS helps here, since there will be a certain degree of spatial separation between $\gamma$-rays coming from the final nucleus and those coming from neutron interactions. This can be exploited to tag $\beta$-delayed neutron events by setting a coincidence gate on the neutron capture "peak" observed, for example, in one half of the spectrometer and looking at the spectra in the other half of the spectrometer. Work to demonstrate the feasibility of this approach is in progress.

\section{Conclusions}

The characterization of the DTAS detector has been carried out. A gain stabilization system based on a light pulse generator has been tested successfully. The non-proportionality of the light yield effects in a NaI(Tl) multi-crystal spectrometer were taken into account to reconstruct properly the sum of the total energy deposited in the spectrometer.
The summing-pileup distortion of the spectrum was successfully 
computed
using a revision of a method previously developed, and for high-rate measurements an improvement in this method has been introduced with the help of MC 
simulated data.
A careful Geant4 MC simulation of the DTAS detector 
response to $\beta$-decays
has been performed.
The quality of the response function, needed for any TAGS analysis, has been validated after obtaining excellent agreement when comparisons were made with measurements of calibration sources. 
This includes in particular a good agreement of multiplicity gated spectra.
A nice agreement between the measured and simulated shape of the $\beta$-delayed neutron background was also obtained
for two well known neutron emitters.

\section{Acknowledgements}
This work has been supported by the Spanish Ministerio de Econom\'ia y Competitividad under grants FPA2011-24553, AIC-A-2011-0696, FPA2014-52823-C2-1-P and the program Severo Ochoa (SEV-2014-0398), by the European Commission under the FP7/EURATOM contract 605203, and by the Spanish Ministerio de Educaci\'on Cultura y Deporte under the FPU12/01527 grant. The work was also supported by the UK Science and Technology Facilities Council (STFC) grant ST/P005314/1. E. Ganio\u{g}lu was supported by the Istanbul University Scientific Research Project Unit under FYO-2017-24144 project.


\bibliography{Guadilla_bibfile2_JLT}

\end{document}